\documentclass[preprint,prc,aps,showpacs,preprintnumbers,eqsecnum,amsmath,amssymb,twocolomn]{revtex4}

\usepackage{graphicx}
\usepackage{epsfig}
\usepackage{dcolumn}
\usepackage{longtable}
\usepackage{bm}

\begin{document}
%
\title{Investigation of
Pygmy Dipole Resonances in the Tin Region}
\author{
 N.~Tsoneva$^{1,2}$, H.~Lenske$^{1}$}
\affiliation{
  $^1$Institut f\"ur Theoretische Physik, Universit\"at Gie\ss en,
  Heinrich-Buff-Ring 16, D-35392 Gie\ss en, Germany \\
$^2${Institute for Nuclear Research and Nuclear Energy, 1784 Sofia, Bulgaria}}

\date{\today}

%
\begin{abstract}
The evolution of the low-energy electromagnetic dipole response with
the neutron excess is investigated along the Sn isotopic chain
within an approach incorporating Hartree-Fock-Bogoljubov (HFB) and multi-phonon
Quasiparticle-Phonon-Model (QPM) theory.
General aspects of the relationship of nuclear skins and dipole sum
rules are discussed. Neutron and proton transition densities serve
to identify the Pygmy Dipole Resonance (PDR) as a generic mode of
excitation. The PDR is distinct from the GDR by its own
characteristic pattern given by a mixture of isoscalar and isovector
components. Results for the $^{100}$Sn-$^{132}$Sn isotopes and the
several N=82 isotones are presented. In the heavy Sn-isotopes the
PDR excitations are closely related to the thickness of the neutron
skin. Approaching $^{100}$Sn a gradual change from a neutron to a
proton skin is found and the character of the PDR is changed
correspondingly. A delicate balance between Coulomb and strong
interaction effects is found. The fragmentation of the PDR strength in $^{124}$Sn is investigated by multi-phonon calculations. Recent measurements of the dipole response in $^{130,132}$Sn are well reproduced.

\end{abstract}

\maketitle

\section{Introduction}

The recent developments in experimental facilities of fast
radioactive beams allows to study exotic nuclei far from the valley
of beta-stability towards the neutron and proton driplines. One of
the most interesting results was the discovery of a new dipole mode
at low excitation energy. Typically, one observes in nuclei with a
neutron excess, $N>Z$, a concentration of electric dipole states at
or close to the particle emission threshold. Since this bunching of
$1^-$ states resembles spectral structures, otherwise known to
indicate resonance phenomena, these states have been named Pygmy
Dipole Resonance (PDR). However, only a tiny fraction of the
Thomas-Reiche-Kuhn energy weighted dipole sum rule strength is found
in the PDR region. Hence, these states will not alter significantly
the conclusions about the importance of the Giant Dipole Resonance
(GDR) in photonuclear reactions, as known for a long time
\cite{Gol48}.

Empirically, the low-energy PDR component increases with the charge
asymmetry of the nucleus. The experimental situation from
high-precision photon scattering experiments performed in
neutron-rich stable nuclei has been reviewed recently in
\cite{PDRrev:06}. Of special interest are as well the newly
performed experiments with radioactive beams in unstable oxygen
\cite{Try,Lei} and tin isotopes, where observation of pygmy dipole
strength has been reported \cite{Adrich:2005}.

In \cite{nadia:2004a,nadia:2004b,Volz:2006} we have established the PDR as a mode
directly related to the size of the neutron skin. By theoretical
reasons the PDR mode should appear also in nuclei with a proton
excess \cite{Paar2005}. Hence, the PDR phenomenon is closely
related to the presence of an excess of either kind of nucleons.

However, considering the measured dipole response functions the
identification of a PDR mode is by no means unambiguous. While in a
nucleus like $^{208}$Pb most of the dipole transition strength is
found in the rather compact GDR region, the picture becomes more
complicated in neutron-rich nuclei between the major shell closures.
The dipole strength shows a tendency to fragment into two or several
groups once the region of stable closed shell nuclei is left
although the GDR still exhausts most of the (almost)
model-independent Thomas-Reiche-Kuhn sum rule. A distinction between
different dipole modes simply by inspection of the nuclear dipole
spectra is clearly insufficient. As we will discuss in later
sections a closer analysis of our theoretical results reveals a
persistence of the typical GDR pattern of an out-of-phase
oscillation of protons and neutrons in most of the low-energy
satellites until a sudden change happens in those parts of the
spectra in the vicinity of the particle emission threshold. As we
have pointed out before \cite{Rye02,nadia:2004a,nadia:2004b} these
states are special in the sense, that they are dominated by
excitation of nucleons of one kind with small admixtures of the
other kind. In a nucleus with neutron excess these excitations
involve mainly neutron particle-hole ($ph$) configurations$^1$.
\footnotetext[1]{A certain amount of proton excitations is required
as a compensation in order to suppress the motion of the nuclear
center-of-mass.}

A clarification about the nature of the low-energy dipole strength
in exotic nuclei can only be expected by the help of theory. With
this paper we intend to contribute to this interesting problem by a
systematic study of the evolution of dipole modes in the Sn
isotopes. As suitable quantities, we consider the non-diagonal
elements of the one-body density matrix. These transition densities
give a snapshot of the motion of protons and neutrons during the
process of an excitation. In principle, they are observables, e.g.
for selected cases as the transition form factors in inelastic
electron scattering \cite{JHeisenberg}. In practice, however,
experimental difficulties are usually inhibiting such measurements.

Considering neutron-rich nuclei the character of the mean-field
changes with increasing neutron excess, because of the enhancement of
the isovector interactions. This has important consequences for the
binding mechanism. In a neutron-rich nucleus the neutron excess
leads to a rather deep effective proton potential, but produces a
very shallow neutron mean-field. This results in deeply bound proton
orbits with separation energies of the order of 20~MeV or even more
as seen in empirical mass tables \cite{Audi95}. The increase in
proton binding is accompanied by a decrease in neutron binding. The
most extreme cases are the spectacular halo states in light nuclei,
e.g. \cite{b8,c19}. However, even for less extreme conditions
unusual nuclear shapes are expected. In neutron-rich medium- and
heavy-mass nuclei neutron skins of a size exceeding the proton
distribution by up to about 1~fm have been predicted. Measurements
in the Na \cite{Na-skin} and Sn \cite{Sn-skin1,Sn-skin2} regions
indeed confirm such conjectures, although the present data are not
yet fully extending into the regions of extreme asymmetry.

The manuscript is organized as follows. In Section
\ref{sec:Features} relations between a neutron or proton skins and
the total nuclear dipole yield are discussed. In section
\ref{sec:HFB} the theoretical methods applied in this paper are
explained. The mean-field part is treated microscopically by using
HFB theory, but for the QPM calculations we allow empirical
adjustments, which we incorporate by an phenomenological Density
Functional Theory (DFT) approach. The description of response functions and
transition densities by QRPA theory is discussed in section
\ref{sec:QRPA}. As an interesting mass region we explore the
unstable tin isotopes and present results on the dipole response in
section \ref{sec:results}. Section \ref{sec:OthersData} is devoted
to comparisons to results of other calculations and data. The paper
closes with a summary and an outlook in section \ref{sec:summary}.

\section{Nuclear Skins and the Dipole Response}\label{sec:Features}
\subsection{The Dipole Excitations in Exotic Nuclei}\label{ssec:ExoticDipole}
In this section we spend a few lines on discussing some relations
between a proton or neutron skin and Pigmy Dipole Modes in exotic
nuclei. We begin with recalling the definition of the nuclear electric dipole operator which in units of the electric charge and in terms of the {\em intrinsic} coordinates $\xi_i$ is given by \cite{Feshbach:V1}
\begin{equation}\label{eq:DipXi}
\vec{D}=\frac{1}{2}\sum_i{\vec{\xi}_i(1-\tau_{3i})}=-\frac{1}{2}\sum_i{\vec{\xi}_i\tau_{3i}} \quad .
\end{equation}
However, in nuclear structure calculations like the present one the particle coordinates $\vec{r}_i$ are used which are related to the intrinsic coordinates by
\begin{equation}\label{eq:Xi2r}
\vec{\xi}_i=\vec{r}_i-\vec{R}
\end{equation}
with the center-of-mass coordinate
\begin{equation}\label{eq:Rcm}
\vec{R}=\frac{1}{A}\sum_i{\vec{r}_i} \quad .
\end{equation}
Be means of eq. \ref{eq:Xi2r} we find
\begin{eqnarray}
\vec{D}&=&-\frac{1}{2}\sum_i{(\tau_{3i}-\frac{T_3}{A})\vec{r}_i}\\
       &=&q_nN\vec{R}_n+q_pZ\vec{R}_p \\
       &=&\frac{NZ}{A}\left(\vec{R}_p-\vec{R}_n\right) \label{eq:RpRn}
\end{eqnarray}
where $T_3$ is the 3-component of the total nuclear isospin operator which is a conserved quantity and can be replaced by its eigenvalue $N-Z$. The neutron and proton recoil corrected effective charges are denoted by $q_n=-Z/A$ and $q_p=N/A$, respectively. The partial sums over proton and neutron coordinates,  normalized to the respective particle numbers, are denoted by $\vec{R}_{p,n}$. They describe the position of the center-of-mass of the protons and neutrons, but neither of the two quantities are conserved  separately.

As is obvious from eq. \ref{eq:DipXi} only the protons participate actively in the radiation process while the neutrons follow their motion such that the position of the center of mass remains unperturbed as is reflected by eq. \ref{eq:RpRn}. This condition implies that $\vec{R}=\frac{N}{A}\vec{R}_n+\frac{Z}{A}\vec{R}_p$ is a stationary operator. In other words, in any nucleus, whether stable or short-lived exotic, the condition $\dot{\vec{R}}\equiv 0$ must be fulfilled, leading to $\dot{\vec{R}}_n=-\frac{Z}{N}\dot{\vec{R}}_p$. Together with eq. \ref{eq:RpRn} this relation reflects the well known property of the dipole giant resonance (GDR) of an oscillation of the proton fluid against the neutron fluid,  also found in hydrodynamical models of the nuclear dipole response \cite{Mohan:1971}.

However, already in \cite{Mohan:1971} it was pointed out that in a nucleus with a neutron excess $N_e=A-2Z>0$ additional modes of excitation are possible. Such a system will also develop modes in which the excess nucleons are oscillating against the bulk, consisting of an equal amount of protons and neutrons, $Z_b\sim N_b$. In that case we write  $\vec{R}=\frac{N_e}{A}\vec{R}_e+\frac{N-N_e}{A}\vec{R}_b+\frac{Z}{A}\vec{R}_p$ and another mode conserving the total center of mass is
\begin{equation}\label{eq:Rmotion}
\dot{\vec{R}}_e=-\left(\frac{N_b}{N_e}\dot{\vec{R}}_b+\frac{Z}{N_e}\dot{\vec{R}}_p\right)
\end{equation}
where $R_b$ denotes the position of the center of mass of the remaining $N_b=N-N_e$ bulk neutrons. Because of $N_b=Z_b=Z$ we find the relation
\begin{equation}\label{eq:SkinMode}
\dot{\vec{R}}_e=-\frac{Z}{N_e}\left(\dot{\vec{R}}_b+\dot{\vec{R}}_p\right) \quad ,
\end{equation}
indicating the motion of the excess neutrons against the core with an equal amount of neutrons and protons. Correspondingly, in a proton-rich nucleus, $Z_e=A-2N>0$, the excess protons may oscillate against the charge-symmetric bulk. In either case, watching that type of motion from the laboratory frame the core neutrons and protons will be seen to move in phase among themselves, but oscillate against the excess component, although the electric dipole operator $\vec{D}$ will couple directly only to the protons\footnotetext[2]{We neglect higher order effects from the non-vanishing electric form factor of the neutron.}.

If we assume harmonic motion, $\vec{R}_{b,p,e}(t)=\vec{R}^{(0)}_{b,p,e}e^{-i\omega t}$, eq.\ref{eq:SkinMode} leads to a relation among the amplitudes
\begin{equation}\label{eq:SkinAmp}
\vec{R}^{(0)}_e=-\frac{Z}{N_e}\left(\vec{R}^{(0)}_b+\vec{R}^{(0)}_p\right) \quad ,
\end{equation}
showing that in a neutron-rich nucleus the skin components will oscillate with an amplitude reduced by the factor $\frac{Z}{N_e}$ compared to the core.

\subsection{Skin Thickness and Dipole Response}\label{ssec:SkinDipole}

The skin thickness is defined usually by the difference of
proton and neutron root-mean-square ($rms$) radii
\begin{equation}\label{eq:skinthick}
\delta r=\sqrt{<r^2_n>}-\sqrt{<r^2_p>},
\end{equation}
where
\begin{equation}\label{eq:rms}
<r^2_q>=\frac{1}{A_q}\int{d^3r r^2 \rho_q(\vec{r})}
\end{equation}
denotes the rms radius of the proton and ($q=p$) and neutron
($q=n$) ground state density distributions $\rho_q$, respectively,
normalized to the corresponding particle number $A_q=N,Z$. For the present purpose a better suited choice is to express the differences in rms-radii in terms of the {\em intrinsic} coordinates, weighted by the 3-component of the isospin operator with eigenvalues $\pm 1$ for neutrons and protons, respectively,
\begin{equation}\label{eq:skinMS}
\Delta_3\xi^2=\langle 0|\sum_i{\xi^2_i\tau_{3i}}|0\rangle=N<\xi^2_n>-Z<\xi^2_p> \quad ,
\end{equation}
which contains the same type of information as eq. \ref{eq:skinthick}.

If we express $\vec{\xi}_i$ in terms of the laboratory coordinates $\vec{r}_i$, eq. \ref{eq:Xi2r}, the ground state expectation value of eq. \ref{eq:skinMS} leads to a corresponding expression in terms of the laboratory coordinates $\{\vec{r}_i\}$,
\begin{equation}
\Delta_3r^2=\sum_i{<0|\tau_{3i}r^2_i|0>}
\end{equation}
\[
=\frac{A}{A-2}\left(\Delta_3\xi^2-\frac{N-Z}{A}<r^2>  \right) \quad .
\]
We have neglected contributions related to two-body correlations.

In terms of the laboratory coordinates the intrinsic nuclear dipole transition operator could be expessed as
\begin{equation}\label{eq:Dproneu}
\vec{D}=\sum_i{\vec{r}_i\left(q_p\frac{1}{2}(1-\tau_{3i})+q_n\frac{1}{2}(1+\tau_{3i})\right)}
\quad .
\end{equation}
Defining the isoscalar ($T=0$) and isovector ($T=1$) charges and space vectors, respectively,
\begin{equation}
q_T=\frac{1}{2}(q_n+(-)^Tq_p)\quad ; \quad \vec{x}_T=\sum_i{\vec{r}_i(\tau_{3i})^T}
\end{equation}
we obtain the isospin representation
\begin{equation}\label{eq:Diso}
\vec{D}=q_0\vec{x}_0+q_1\vec{x}_1 \quad .
\end{equation}
The reduced isovector/isoscalar dipole transition moments 
and the dipole transition probabilities can be expressed as

$\vec{M}^{(T)}_d=<0||(\tau_3)^T\vec{r}||d>$;

\begin{equation}
B_d(E1)=|q_0\vec{M}^{(0)}_d+q_1\vec{M}^{(1)}_d|^2 \quad .
\end{equation}
For the present purpose we are interested in the isoscalar-isovector interference term, in particular
\begin{equation}\label{eq:M0M1}
\Re\sum_d{\vec{M}^{(0)}_d\cdot\vec{M}^{(1)*}_d}=
\end{equation}
\[
=\frac{1}{2q_0q_1}\left(\sum_d{B_d(E1)}-q^2_0\sum_d{|M^{(0)}_d|^2}-q^2_1\sum_d{|M^{(1)}_d|^2}  \right) \quad .
\]
Introducing a single particle basis $\{\varphi(\vec{r})_i \}$ enables us to express the nuclear dipole eigenstates $|d>$ in terms of particle-hole excitations $|\alpha>$.

In the concrete case considered here, we use a QRPA description in terms of two-quasiparticle excitations $|\alpha>=|(ij)JM>$ with the single quasiparticle states $i$ and $j$, respectively, coupled to total angular momentum $JM$. Neglecting ground state correlations, which are of at least $2p2h$ character, the left hand side of eq. \ref{eq:M0M1} can expressed in terms of single particle matrix elements
\begin{equation}
\frac{1}{2}\sum_d{\vec{M}^{(0)}_d\cdot\vec{M}^{(1)*}_d}=\sum_{i,j}{\vec{M}^{(0)}_{ij}\cdot\vec{M}^{(1)*}_{ij}v^2_i}
\end{equation}
\[
-\sum_{i,j}{\vec{M}^{(0)}_{ij}\cdot\vec{M}^{(1)*}_{ij}v^2_iv^2_j}\nonumber \\
+\sum_{i,j}{\vec{M}^{(0)}_{ij}\cdot\vec{M}^{(1)*}_{ij}u_iv_iu_jv_j} \quad 
\]

where $v_i^2=\left\langle \left|a^{+}_{i}a_{i} \right|\right\rangle$ and 
$u_i^2=1-v_i^2$ are occupation numbers.  
The last two terms in the above equation are of 2-body character and appear because of the pairing ground state correlations. For our purpose the first term on the right hand side is of special interest. It is seen to correspond to
\begin{equation}
\Re\sum_{i,j}{\vec{M}^{(0)}_{ij}\cdot\vec{M}^{(1)*}_{ij}v^2_i}=\sum_{i,j}{<i|\vec{r}|j>\cdot<j|\tau_3\vec{r}|i>v^2_i}
\end{equation}
\[
=\sum_{i}{<i|\vec{r}\cdot\tau_3\vec{r}|i>v^2_i}=\sum_{i}{<i|\tau_3r^2|i>v^2_i},
\]
where we have used the completeness of the single particle states.
Hence, we have derived an important theoretical relation between the non-energy weighted dipole sum rule and the skin measure defined before
\begin{equation}\label{eq:skinDipole}
\Delta_3r^2=\frac{1}{4q_0q_1}\left(\sum_d{B_d(E1)}-q^2_0\sum_d{|M^{(0)}_d|^2}\right.
\end{equation}
\[\left.
-q^2_1\sum_d{|M^{(1)}_d|^2}\right)
+\Re\sum_{i,j}{\vec{M}^{(0)}_{ij}\cdot\vec{M}^{(1)*}_{ij}v^2_iv^2_j}
\]
\[
-\Re\sum_{i,j}{\vec{M}^{(0)}_{ij}\cdot\vec{M}^{(1)*}_{ij}u_iv_iu_jv_j} \quad .
\]
In the first term the pure isovector and isoscalar dipole sum rule strengths are subtracted off the full dipole sum rule, thus leaving the interference term. The isovector sum rule will be dominated, if not exhausted, by the GDR. From the Thomas-Reiche-Kuhn sum rule, i.e. the corresponding energy weighted dipole sum rule, we know that the GDR strength varies little along an isotopic chain, namely as $NZ/A$. With $N=N_b+\delta N$,$A=A_b+\delta N$, and $Z_b=Z$ we find $\frac{NZ}{A}\sim\frac{N_bZ_b}{A_b}(1+\frac{\delta N}{A_bN_b})$. We emphasize again that the isoscalar sum rule is a pure recoil effect, appearing only in the neutron-rich nuclei and expressing the compensating motion of the neutrons in the laboratory frame.

These relations reveal the intimate connection between the neutron skin (which is a static property) and the dipole spectrum (which is a dynamical property): From the above equations we find that the skin thickness is directly related to the dipole response. By eq. \ref{eq:skinDipole} another aspect is emphasized, namely the fact that, seen from the laboratory, apparent isoscalar and isovector moments seem to contribute to the excitation of dipole states. As discussed above, modes involving an isoscalar component are allowed provided that the position $\vec{R}$ of the total nuclear center of mass is left untouched.

\subsection{Dipole Excitations and Spurious States}

The problem posed by broken symmetries in effective nuclear Hamiltonians like ours is well known, e.g. \cite{Thouless:61,Meyer:79}. For dipole excitations the most relevant effects are due to the violation of translational and Galilean symmetry. A known property of RPA is to restore the broken
translational symmetry by generating a states at zero excitation
energy, corresponding to a symmetry-restoring {\em Goldstone-mode}, provided that a complete configuration space was used. The transition strength scales with the total particle number $A=N+Z$ and exhaust to a large extent the isoscalar sum rule. In practice, that is hardly achieved. But numerically we can enforce the restoration by a proper choice of the residual interaction in the isoscalar dipole channel. An alternative are projection techniques which, to our knowledge have never been applied to a realistic multi-configuration QRPA
calculation.

In this paper, we are considering the following situation. In a $N\gg Z$ nucleus the conditions change insofar as new {\em intrinsic} excitations (see eq.(\ref{eq:SkinMode})) will
appear, not encountered in stable $N\simeq Z$ nuclei. As already
pointed out some time ago by Mohan {\em et al.} \cite{Mohan:1971} in
a neutron-rich nucleus the excess neutrons may be excited into
oscillations against the core, either in phase or out of phase with
the core protons. Especially the latter mode is the one from which
we can expect a sizable content of isoscalar strength. That mode,
however, will never appear as a pure isoscalar mode because of a
compensating motion of the core neutrons required in order to keep
fixed the center-of-mass of the whole system. Obviously, this is an
intrinsic mode which will be strongly suppressed when approaching
the $N=Z$ line. In fact, the isoscalar content of the PDR states is
impressively confirmed by a recent experiment in $^{140}$Ce
\cite{Savran:2006}, comparing spectra from inelastic scattering of
$\alpha$ particles to $(\gamma,\gamma')$ spectra. Since the $\alpha$
particle is a pure isospin $T=0$ probe it acts as an isospin filter
and the spectra in \cite{Savran:2006} show clearly the content of
isoscalar transition strength in the PDR region. The special
character of these transitions becomes clear from the shapes of the
transition densities shown later which obviously do not resemble any
expectations from classical or semi-classical models.

The GDR mode as one of the most collective excitations in nuclei is
well understood, both quantum mechanically and in semi-classical
hydrodynamical approaches, while the nature of the PDR is still
waiting for full clarification. The afore mentioned early attempts
to incorporate the low-energy dipole modes into the hydrodynamical
scenario \cite{Mohan:1971} by a three-fluid {\em ansatz} seemed to
work reasonably well in $^{208}$Pb, but when applied to the
Ca-isotopes \cite{Suzuki:1990} the model failed as pointed out by
Chambers et al. \cite{Cha94}. Our more detailed microscopic QPM
studies of the PDR strength in $^{208}$Pb \cite{Rye02}, including
transition densities and currents, gave strong indications, that the
PDR modes are of generic character, clearly distinguishable from the
established interpretation of the GDR by strong vorticity
components. The differences are also visible in the transition
densities, where they are showing up in terms of a nodal structure,
unknown from GDR excitations. Hence, we have the surprising
situation that a mode with a more complex spatial pattern is seen at
energies below the most collective state. This (theoretical)
observation indicates that PDR and GDR states are indeed belonging
to distinct parts of the nuclear spectrum. The situation is less
confusing if we take the view that the PDR is related to a more
complex excitation scenario as indicated by the transition densities
and velocity fields discussed in \cite{Rye02}. The characteristic
features of PDR transition densities will be investigated in the
following for the whole chain of known Sn isotope, from $^{100}$Sn
to $^{132}$Sn.

\section{Phenomenological Density Functional Approach for Nuclear Ground States}\label{sec:HFB}

\subsection{The Density Functional}\label{ssec:DFT}
Our method is based on a fully microscopic HFB
description of the nuclear ground states and the quasiparticle
spectra as the appropriate starting point for a single-phonon QRPA
or multi-phonon QPM calculation of nuclear spectra. However, being
aware of the deficiencies of existing density functionals, when
leaving the region of stable nuclei we accept slight
adjustments and phenomenologically motivated choices of parameters.
We assure a good description of nuclear ground state properties by
enforcing that measured separation energies and nuclear radii are
reproduced as close as possible.

We start by considering the ground state of an even-even nucleus in
an independent quasiparticle model for which we use the microscopic
HFB approach. The nucleons move in a static mean-field, which is
generated self-consistently by their mutual interactions including a
monopole pairing interaction in the particle-particle (pp) channel.
Following the DME approach of \cite{Hofmann,Hofmann01}, the
interactions are taken from a G-Matrix, but renormalized such that
nuclear matter properties are reproduced, thus accounting
effectively for correlations missed by a static two-body
interaction. In local density approximation the problem is then
reduced essentially to the level of a Skyrme-HFB calculation as
discussed in \cite{Hofmann}. An important difference, however, is
the use of a microscopically obtained density dependent pairing
interaction. The HFB and BCS equations are solved self-consistently
with state dependent gaps by iteration, until convergence of the
mean-field and single-particle energies, gaps and densities is
achieved.

The single-particle energies and ground state properties in general
are critical quantities for extrapolations of QRPA and QPM
calculations into unknown mass regions. Here, we put special
emphasis on a reliable description of the mean-field part, reproducing as close as possible the g.s. properties of nuclei along an isotopic chain. This is achieved by solving the ground state problem in a semi-microscopic approach.
Following the arguments given in \cite{nadia:2004a} we take
advantage of the Hohenberg-Kohn \cite{HoKohn:64} and Kohn-Sham
\cite{KohnSham:65} theorems, respectively, of density functional
theory, which state, that the total binding energy $B(A)$ can always
be expressed as an integral over an energy density functional with a
(quantal) kinetic energy density $\tau$ and density dependent
self-energy parts $U(\rho)$, respectively,
\begin{equation}\label{eq:EDF}
B(A)=\sum_{q=p,n}{\int{d^3r\left( \tau_q(\rho)+\frac{1}{2}\rho_q
U_q(\rho)\right)}+E^{pair}_q} \quad ,
\end{equation}
where we have chosen a representation in terms of proton ($q=p$) and
neutron ($q=n$) densities $\rho_q=\rho_q(\vec{r})$, respectively, as
appropriate for nuclei far from the stability region with exotic charge-to-mass
ratios. The total isoscalar $(T=0)$ and isovector $(T=1)$ densities
are defined by $\rho_{T}=\rho_n+(-)^T\rho_p$
and $\rho_0$ is normalized to the total particle number A.

In addition to the kinetic and potential energy terms eq.
(\ref{eq:EDF}) includes pairing contributions, which are indicated
separately by $E^{pair}_q$. In fact, this means, that we use an
extended version of the Kohn-Sham theorem including the proton and
neutron pairing densities $\kappa_q$ as well, as dictated by HFB
theory. Hence, the density functional underlying in eq. (\ref{eq:EDF}) is
of the form $\mathcal{E}(\tau,\rho,\kappa)$, where each of the
arguments are understood to include proton and neutron parts,
respectively.

In terms of the single-particle wave functions
$\varphi_{jq}(\vec{r})$ and the occupancies $v^2_{jq}$ the kinetic
energy density is given by
\begin{equation}\label{eq:KinE}
\tau_q=\sum_{j}{v^2_{jq}\frac{\hbar^2}{2M_q}|\vec{\nabla}\varphi_{jq}(\vec{r})|^2}
\end{equation}
and the number and pairing densities are
\begin{eqnarray}
\rho_q(\vec{r})&=&\sum_j{v^2_{jq}|\varphi_{jq}(\vec{r})|^2} \label{eq:densities}\\
\kappa_q(\vec{r})&=&\frac{1}{2}\sum_j{v_{jq}u_{jq}|\varphi_{jq}(\vec{r})|^2},
\label{eq:kappa}
\end{eqnarray}
where $v_{jq},u_{jq}$ denote BCS amplitudes with
$u^2_{jq}=1-v^2_{jq}$. The summations over $j$ includes the full set
of quantum numbers specifying the single-particle states
$\varphi_{jq}(\vec{r})$.

Rather than using a conventional density functional like the Skyrme
functional we choose to express the interaction part in terms of a
superposition of central and spin-orbit potentials of Wood-Saxon
shape. This {\em ansatz} gives us the full flexibility to describe
nuclear ground state properties like binding energies, root mean
square radii, and separation energies to the required accuracy. The
price to be paid is a lack of contact to a fully microscopic picture
like in \cite{Hofmann}. However, for the present purpose and in view
of the persisting uncertainties on the dynamics in strongly
asymmetric nuclear matter, we are convinced, that a phenomenological
approach allowing a self-consistent description of nuclear ground
states is an eligible method.

Hence, in order to describe the bulk properties of the nuclear
ground states in the best possible manner, we decide to be satisfied
by using functionals optimized to a given mass region, in this case
the Sn isotopes. The parameters of the model are the strengths, the
radii and the diffuseness parameters of the corresponding
parameters. {\em A posteriori} the collected information will allow
us to derive eventually a nuclear energy density functional of
general applicability. In other words, we try to avoid a biased
choice of operators by assuming a certain operator structure at the
level of a two-body interaction.

\subsection{The Single-Particle States}\label{ssec:spStates}

From eq. (\ref{eq:EDF}) we derive by variation a Schroedinger
equation
\begin{equation}\label{eq:WaveEq}
\left(-\frac{\hbar^2}{2M_q}\vec{\nabla}^2+\Sigma_q(\vec{r})-\eta_{jq}\right)\varphi(\vec{r})=0
\end{equation}
for the single-particle wave functions $\varphi_{jq}$ and
eigen energies $\eta_{jq}$. The self-energy $\Sigma_q$ appearing in
eq. (\ref{eq:WaveEq}) is obtained variationally from the interaction
energy density
\begin{equation}\label{eq:Eint}
E_{int}=\frac{1}{2}\sum_{q}{\rho_{q}U_{q}(\rho)},
\end{equation}
where we have defined the single-particle occupation probabilities
$\rho_q$, which in BCS approximation are given by $v^2_q$. 
By variation with
respect to $\rho_q$ we obtain
\begin{equation}\label{eq:defSelfE}
\Sigma_q(\rho)=\frac{1}{2}\frac{\partial}{\partial
\rho_q}_{|v^2_q}\sum_{q'}{\rho_{q'}U_{q'}(\rho)} \quad .
\end{equation}
Because of the intrinsic density dependence of $U_q(\rho)$ we find,
that $\Sigma_q$ differs from the proper interaction energy by a
rearrangement potential
\begin{equation}\label{eq:SelfE}
\Sigma_q(\rho)=U_q(\rho)+U^{(r)}_q(\rho)
\end{equation}
given by
\begin{equation}\label{eq:Urear}
U^{(r)}_q(\rho)=\frac{1}{2}\sum_{q'}{\left(\rho_{q'}\frac{\partial}{\partial
\rho_q}_{v^2_q} U_{q'}(\rho)-\delta_{qq'}U_q(\rho)\right)} \quad .
\end{equation}

which is discussed in more detail in App. \ref{app:A}.
In nuclei with non-vanishing pairing additional contributions from
the density gradients of $E^{(pair)}_q=E^{(pair)}_q(\kappa,\rho)$
will also contribute.

\subsection{Pairing and Quasiparticle States}\label{ssec:PairQP}

From the density functional, eq. (\ref{eq:EDF}), we obtain the proton
and neutron pairing fields $\Delta_q(\rho,\kappa)$ by variation with
respect to the pairing (or anomalous) densities $\kappa_q$, eq.
(\ref{eq:kappa})
\begin{equation}\label{eq:Delta}
\Delta(\rho,\kappa)=\frac{\delta B(A)}{\delta \kappa_q}=\kappa_q
V^{(pair)}(\rho),
\end{equation}
which we decide to factorize into the anomalous density and a local
density dependent pairing strength $V^{(pair)}(\rho)$, depending on
the local bulk density $\rho=\rho(r)$. $V^{(pair)}(\rho)$ is
discussed below.

With the usual Bogolubov transformations we obtain the quasiparticle
states $\alpha^+_{jq}=u_{jq}a^+_{jq}-v_{jq}\tilde{a}_{jq}$
\cite{Sol}. Together with the Schroedinger equation
(\ref{eq:WaveEq}), we solve self-consistently the BCS gap equation for
the state dependent pairing gaps $\Delta_{jq}=\langle
jq|\Delta_q|jq\rangle$ for protons and neutrons, respectively,
\begin{equation}\label{eq:GapEq}
\Delta_{jq}=\sum_{k}{v_{kq}u_{kq}\langle
j\tilde{j}|V^{(pair)}|k\tilde{k}\rangle}=\frac{1}{2}\sum_{k}{\frac{\Delta_{kq}}{E_{kq}}\langle
j\tilde{j}|V^{(pair)}|k\tilde{k}\rangle} \quad .
\end{equation}
Time-reversed states are denoted by a tilde. In a spherical
symmetric nucleus the BCS state amplitudes are
\begin{equation}\label{eq:BCSAmp}
v^2_{jq}=\frac{1}{2}\left(1-\frac{\eta_{jq}-\lambda_q}{E_{jq}}\right)
\end{equation}
with the quasiparticle energy
\begin{equation}\label{eq:BCSEnergy}
E_{jq}=\sqrt{(\eta_{jq}-\lambda_q)^2+\Delta^2_{jq}}
\end{equation}
The pure mean-field single-particle energies $\eta_{jq}$ are
obtained from eq. (\ref{eq:WaveEq}). The proton and neutron chemical
potentials are denoted by $\lambda_{p,n}$, respectively.

In the practical HFB calculation we use a pairing strength of a
simple form
\begin{equation}\label{eq:Vpair}
V^{(pair)}(\rho)=\left(V_{ext}(1-(\frac{\rho}{\rho_0})^\beta)+V_{int}(\frac{\rho}{\rho_0})^\beta\right)
\quad ,
\end{equation}
simulating the on-shell singlet-even NN interaction amplitude and
depending on the local nuclear density $\rho=\rho(\vec{r})$. The
interaction strength $V_{ext}=-9280MeVfm^3$ is determined such, that
asymptotically for $\rho\to 0$ the nucleon-nucleon scattering length
$a_{pp}\sim a_{nn}\sim a_{SE}=-17.3fm$ in the singlet even channel
for like particles is reproduced. $V_{int}=-0.721MeVfm^3$ is fixed
by requiring, that the pairing gap has a maximum of
$\Delta(\rho_c)=2$~MeV at $\rho_c = 1/3\rho_0$ of the equilibrium
density of infinite nuclear matter, $\rho_0=0.16fm^{-3}$. The best
results are obtained for a small value of the density exponent
$\beta=\frac{1}{85}\sim 0.012$. In the pairing calculations we
include proton and neutron single particle states up to the
respective continuum thresholds. In this way, we avoid instabilities
in the BCS equations and the calculations of number and pairing
densities due to the possible admixture of unbound quasiparticle
orbitals into the bound state region. Such an approach is
permissible because in all the consider nuclei the driplines are not
reached, Hence, the more involved treatment by explicitly solving
the coupled Gorkov-equations, discussed e.g. in \cite{RPPN:01}, can
be avoided without a significant loss of accuracy.

\subsection{HFB Results for Sn Isotopes}

As discussed above, we decide to express the full proton and neutron
self-energies $\Sigma_{p,n}$ in terms of (a superposition of)
Wood-Saxon potentials $\Sigma^{(WS)}_{p,n}$ by a least-square fit of
the depth, radius and diffusivity parameters to separation energies
and charge radii, taken either -- if available -- from empirical
mass compilations \cite{Audi95}, or from our HFB calculations. 
Different to the usual HFB approach the s.p. wave equations are 
solved with effective mass m*=m thus removing
the known problem of unrealistically large HFB level spacings at
the Fermi surface. 

\begin{figure*}
\includegraphics[width=13cm,height=9cm,angle=0]{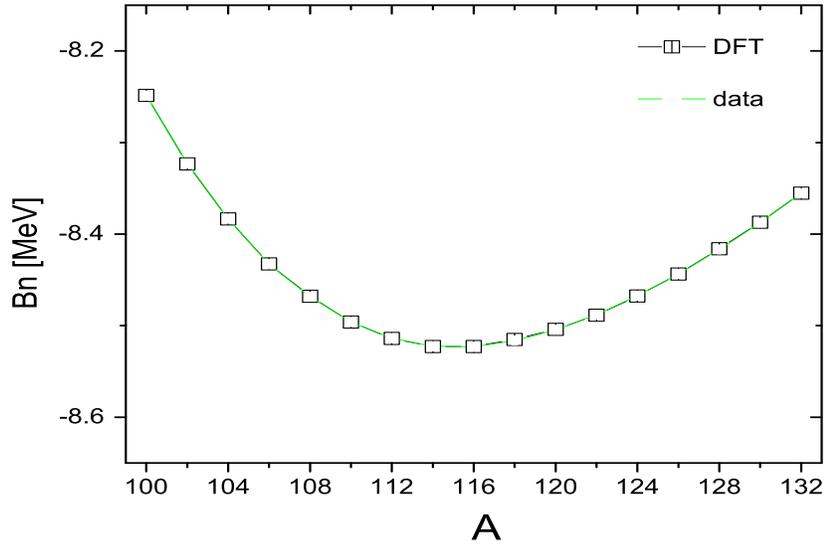}\\
\caption{\label{FIG:FIG1} (Color online) Ground state properties of the Sn
isotopes. The nuclear binding energies per particle calculated with
the DFT approach, discussed in the text, are compared
to data from the Audi-Wapstra compilation \protect\cite{Audi95}.}
\end{figure*}

\begin{center}
\begin{figure*}
\includegraphics[width=17cm,angle=0]{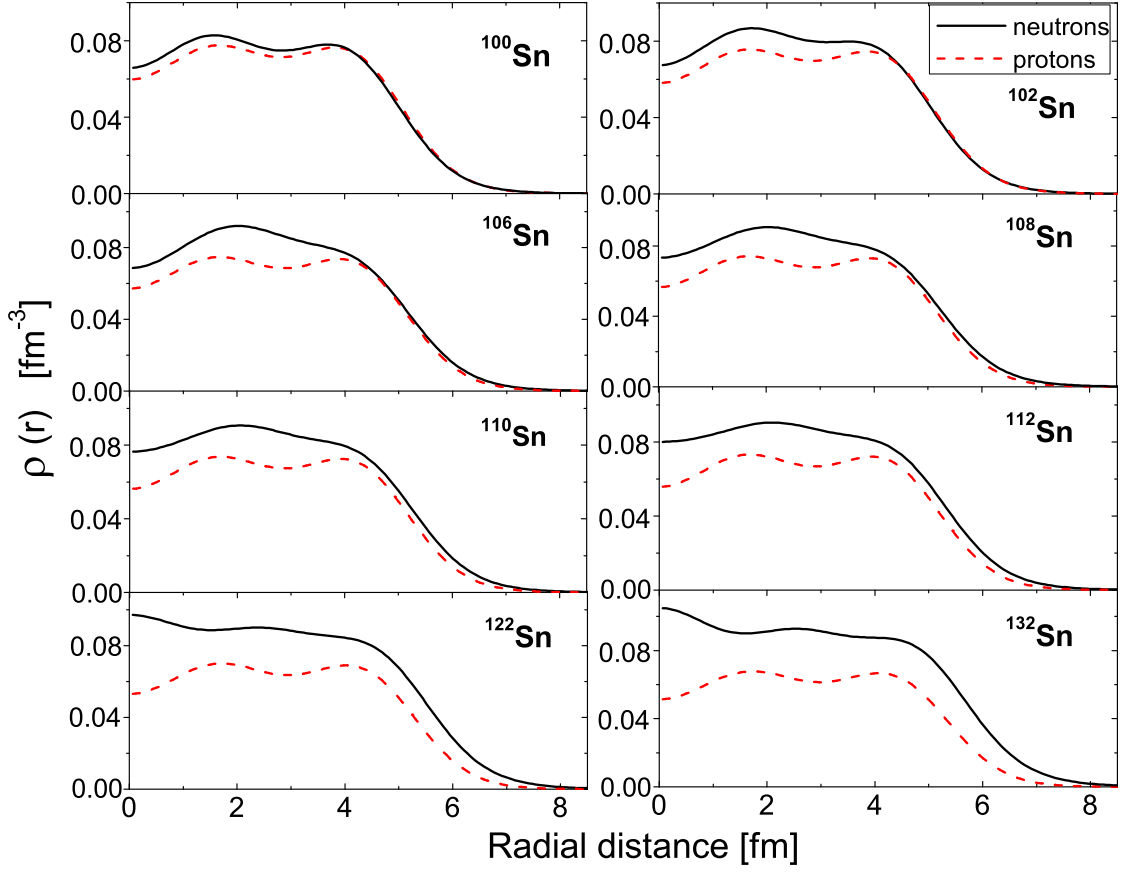}\\
\caption{(Color online) BCS ground state densities of Sn isotopes obtained by the
phenomenological DFT approach and used in the QPM
calculations.} \label{FIG:FIG2}
\end{figure*}
\end{center}

The reproduction of the total binding energy $B(N,Z)$, calculated as
indicated above, of the charge radius and the (relative) differences
of proton and neutron root-mean-square (RMS) radii $\delta r$, taken
from our previous HFB calculations \cite{Hofmann,Hofmann01}, are
imposed as additional constraints. The results of the represent
approach are displayed and compared to measured values in Fig.
\ref{FIG:FIG1}.
The ground state neutron and proton densities are displyed in
Fig.\ref{FIG:FIG2} for several tin isotopes. The comparison between
the neutron and proton densities, obtained by HF calculations with
the D3Y G-matrix interaction (see Fig.7 from ref.\cite{Hofmann}) and
the present ones is very reasonable.

Of special importance for our investigation are the surface regions,
where the formation of a skin takes place as is visible in
Fig.\ref{FIG:FIG2}. For A$\geq$106 the neutron distributions begin
to extend beyond the proton density and the effect continues to
increase with the neutron excess, up to $^{132}Sn$. Thus, these
nuclei have a neutron skin. The situation reverses in $^{100-102}Sn$,
where a tiny proton skin appears.

\section{QPM Description of Nuclear Excitations}\label{sec:QRPA}

\subsection{The QPM Hamiltonian}\label{ssec:QPM}

The excitations are calculated in the framework of the
Quasiparticle-Phonon Model (QPM) with the model Hamiltonian
\cite{Sol}:
\begin{equation}
{H=H_{MF}+H_M^{ph}+H_{SM}^{ph}+H_M^{pp}} \quad .\label{hh}
\end{equation}
Here, $H_{MF}=H_{sp}+H_{pair}$ is the mean-field part, discussed in
the previous section. Hence, different from the standard QPM scheme
we use single-particle energies and wave functions, obtained
self-consistently, according to the procedure described above. For
the QPM calculations the pairing part is simplified by using a
constant matrix element. The method we have applied for the
determination of the ground state properties has been successfully
applied for the investigation of low-lying dipole modes in the tin
isotopic chain before \cite{nadia:2004a,nadia:2004b} and more
recently also in the $N=82$ isotones \cite{Volz:2006}.

$H_M^{ph}$, $H_{SM}^{ph}$ and $H_M^{pp}$ are residual interactions,
taken as a sum of isoscalar and isovector separable multipole and
spin-multipole interactions in the particle-hole and multipole
pairing interaction in the particle-particle channels. The latter is
included only for the quadrupole and octupole excitations.

Building blocks of the model space are the
Quasiparticle-Random-Phase-Approximation (QRPA) phonons:
\begin{equation}
Q^{+}_{\lambda \mu i}=\frac{1}{2}{
\sum_{j_1j_2}{ \left(\psi_{j_1j_2}^{\lambda\mu i}A^+_{\lambda\mu}(j_1j_2)
-\varphi_{j_1j_2}^{\lambda \mu i}\widetilde{A}_{\lambda\mu}(j_1j_2)
\right)}} \label{eq:StateOp}
\end{equation}
defined as a linear combination of two-quasiparticle creation
${A}^+_{\lambda \mu}$ and annihilation operators
$\widetilde{A}_{\lambda \mu}$, respectively. The latter is the time
reversed operator $\widetilde{A}_{\lambda \mu}=(-)^{\lambda
-\mu}A_{\lambda-\mu}$.

Here $j\equiv{(nljm\tau)}$ is a single-particle proton or neutron state.

The (bare) two-quasiparticle operators
\begin{equation}
A^+_{\lambda \mu}(j_1j_2)=\left[\alpha^+_{j_1 }\alpha^+_{j_2 }\right]_{\lambda\mu}
\end{equation}
are defined by coupling the one-quasiparticle operators to total
angular momentum $\lambda$ with projection $\mu$
\begin{equation}
[\alpha^{+}_{j_1} \alpha^{+}_{j_2}]_{\lambda
\mu}=\sum_{m_1m_2}C^{\lambda\mu}_{j_1m_1j_2m_2}\alpha^{+}_{j_1 m_1} \alpha^{+}_{j_2 m_2}
\label{clebsch}
\end{equation}
by means of the Clebsch-Gordan coefficients
$C^{\lambda\mu}_{j_1m_1j_2m_2}=\left\langle
j_1m_1j_2m_2|\lambda\mu\right\rangle$.

The QRPA states are normalized according to the condition
\begin{equation}
\langle 0|Q_{\lambda \mu i}Q^+_{\lambda \mu
i}|0\rangle=1,
\end{equation}
which can be rewritten in terms of two-quasiparticle weight factors  
\begin{equation}
\sum_{j_1>j_2}{w_{j_1j_2}(\lambda\mu i)}=1
\end{equation}

\[
w_{j_1j_2}({\lambda \mu i})=|\psi_{j_1j_2}^{\lambda\mu
i}|^2-|\varphi_{j_1j_2}^{\lambda\mu i}|^2
\]

The weight factors $w_{j_1j_2}({\lambda \mu i})$
are given for some states in Table I and Table II,
respectively.

The QRPA operators obey the equation of motion
\begin{equation}\label{eq:EoM}
\left[H,Q^+_\alpha\right]=E_\alpha Q^+_\alpha \quad ,
\end{equation}
which solves the eigenvalue problem, giving the excitation energies
$E_\alpha$ and the time-forward and time-backward amplitudes
\cite{Sol} $\psi_{j_1j_2}^{\lambda i}$ and $\varphi_{j_1j_2}^{\lambda i}$,
respectively.

The QPM Hamiltonian (\ref{hh}) is rewritten in terms of phonons \cite{Sol}:
\begin{eqnarray}
 H=H_{ph} + H_{qph}=
\sum_{\lambda \mu i}E _{\lambda i}Q_{\lambda \mu i}^{+}Q_{\lambda
\mu i}^{}
\label{hp}
\end{eqnarray}
\begin{eqnarray}
+\frac{1}{2}\sum_{{\lambda _1\lambda _2\lambda _3}{{
{i_1i_2i_3}{\mu _1\mu _2\mu _3}}}}C_{\lambda _1\mu _1\lambda _2\mu
_2}^{\lambda _3{-\mu _3}}
U_{\lambda _1i_1}^{\lambda _2i_2}({\lambda _3i_3})
\nonumber
\end{eqnarray}
\begin{eqnarray}
[Q_{\lambda _1\mu
_1i_1}^{+}Q_{\lambda _2\mu _2i_2}^{+}Q_{\lambda _3-\mu
_3i_3}^{}+h.c.]
\nonumber
\end{eqnarray}

The first term in the eq.(\ref{hp}) contains free phonon operators and refers to the harmonic part of
nuclear vibrations, while the second one is accounting for the
interaction between quasiparticles and phonons. The latter reflect in anharmonic effects and fragmentation of the nuclear excitations.

The Hamiltonian (\ref{hp}) is diagonalized assuming a spherical $0^+$
ground state which leads to an orthonormal set of wave functions
with good total angular momentum JM. For even-even nuclei these
wave functions are a mixture of one-, two- and three-phonon
components \cite{Gri} in the following way:

\begin{equation}
\Psi_{\nu} (JM) =
 \left\{ \sum_i R_i(J\nu) Q^{+}_{JMi}
\right.
\label{wf}
\end{equation}
\[
\left.
+ \sum_{\scriptstyle \lambda_1 i_1 \atop \scriptstyle \lambda_2 i_2}
P_{\lambda_2 i_2}^{\lambda_1 i_1}(J \nu)
\left[ Q^{+}_{\lambda_1 \mu_1 i_1} \times Q^{+}_{\lambda_2 \mu_2 i_2}
\right]_{JM}
{+ \sum_{_{ \lambda_1 i_1 \lambda_2 i_2 \atop
 \lambda_3 i_3 I}}}
\right.
\]
\[
\left.
{T_{\lambda_3 i_3}^{\lambda_1 i_1 \lambda_2 i_2I}(J\nu )
\left[ \left[ Q^{+}_{\lambda_1 \mu_1 i_1} \otimes Q^{+}_{\lambda_2 \mu_2
i_2} \right]_{IK}
\otimes Q^{+}_{\lambda_3 \mu_3 i_3}\right]}_{JM}\right\}\Psi_0
\]
where R,P and T are unknown amplitudes, and $\nu$ labels the
number of the  excited states.

The nuclear response on an external electromagnetic field is
described in terms of quasiparticles and phonons by a transition
operator composed of two parts:
\begin{equation}
M(\mbox{E(M)} \lambda \mu ) = M^{ph}(\mbox{E(M)} \lambda \mu )+
M^{qph}(\mbox{E(M)} \lambda \mu )
\label{EM}
\end{equation}
The first part is responsible for the transitions with one-phonon
exchange between the initial and final states. The
second one contains structures
$[\alpha^{+}_{j}\otimes\alpha_{j'}]_{\lambda \mu}$ including the
interaction between quasiparticles and phonons. It is important
for the description of the so-called {\it boson forbidden}
transitions between nuclear states with the same number of phonons
or differing by an even number of them. The corresponding equations of each of the terms could be found in \cite{Pon}.

\subsection{Transition Densities}\label{ssec:TRD}
In order to understand the character of a nuclear excitation it is
useful to consider the spatial structure of the transition. This is
accomplished by analyzing the one-body transition densities
$\delta\rho (\vec{r})$, which are the non-diagonal elements of the
nuclear one-body density matrix. Physically, $\delta\rho (\vec{r})$
corresponds to the density fluctuations, induced by the action of an
(external) one-body operator on the nucleus. Hence, the transition
densities are directly related to the nuclear response functions and
by analyzing their spatial pattern we obtain a very detailed picture
of e.g. the radial distribution and localization of the excitation
process. The particular usefulness of such an analysis for PDR
states was pointed out in \cite{Rye02}.

Using the complete set of single-particle states
$\varphi_{j}(\vec{r})$ from $H_{MF}$ and a multipole expansion by
means of the Wigner-Eckardt theorem, we find the isoscalar (T=0) and isovector (T=1) transition densities in second quantization:
\begin{equation}\label{eq:TRD}
\delta\rho^T(\vec{r})=\sum_{j_1j_2;\lambda \mu }{\left(i^\lambda Y_{\lambda \mu}(\hat{r})\right)^\dag\rho^{\lambda T}_{j_1j_2}(r)
\left[a^+_{j_1}a_{j_2}\right]_{\lambda \mu}}
\quad .
\end{equation}

For the present purpose we consider non-spin flip transitions of
isoscalar and isovector character. The radial parts are given by
binomials of radial single-particle wave functions and reduced
matrix elements
\begin{equation}
\rho^{\lambda T}_{j_1j_2}(r)=R^*_{j_1}(r)R_{j_2}(r)\frac{1}{\hat{\lambda}}\langle
j_1||i^\lambda Y_\lambda||j_2\rangle\langle q|\tau^T_3|q\rangle \quad ,
\end{equation}
with $\hat{\lambda}=\sqrt{2\lambda+1}$.
The isospin matrix element $\langle q|\tau^T_3|q\rangle$ is unity for T=0. For an isovector transition we have $\langle q|\tau^T_3|q\rangle=\pm 1$ for neutrons and protons, respectively.

The transition densities are obtained by the matrix elements between
the ground state $|\Psi_i\rangle=|J_iM_i\rangle$ and the excited
states $|\Psi_f\rangle=|J_fM_f\rangle$,
\begin{equation}
\rho_{if}^T(\vec{r})=\sum_{j_1j_2;\lambda \mu}{\left(i^\lambda Y_{\lambda \mu}(\hat{r})\right)^\dag\rho^{\lambda T}_{j_1j_2}(r)} \langle
J_jM_f|\left[a^+_{j_1}a_{j_2}\right]_{\lambda \mu}|J_iM_i\rangle
\end{equation}
Here, we are interested only in the two-quasiparticle creation and
annihilation parts which are given by
\begin{equation}
\Gamma^+_{\lambda\mu}(j_1j_2)=\left(u_{j_1}v_{j_2}+v_{j_1}u_{j_2}\right)\left(A^+_{\lambda\mu}(j_1j_2)+\widetilde{A}_{\lambda \mu}(j_1j_2)\right)
\label{eq:gammaA}
\end{equation}
Equation (\ref{eq:gammaA}) can be rewritten in terms of QRPA phonons
defined by the relation (\ref{eq:StateOp}):
\begin{equation}
\Gamma^+_{\lambda\mu}(j_1j_2)=\sum_i{g^{\lambda i}_{j_1j_2}
\left(Q^+_{\lambda\mu i}+\widetilde{Q}_{\lambda \mu
i}\right)}, \label{gammaA}
\end{equation}
where
\begin{equation}\label{eq:gFac}
g^{\lambda i}_{j_1j_2}=\frac{\psi_{j_1j_2}^{\lambda
i}+\varphi_{j_1j_2}^{\lambda i}}{1+\delta_{j_1j_2}}\left(u_{j_1}
v_{j_2}+u_{j_2}v_{j_1}\right)
\end{equation}
accounts for the BCS and QRPA properties, respectively. Thus, we
find
\begin{equation}
\rho_{if}^T(\vec{r})=\sum_{j_1j_2;\lambda\mu}{\left(i^{\lambda}Y_{\lambda\mu}(\hat{r})\right)^\dag\rho^{\lambda T}_{j_1j_2}(r)} \langle
J_fM_f|\Gamma^+_{\lambda\mu}(j_1j_2)|J_iM_i\rangle \quad .
\end{equation}
We identify $|J_iM_i\rangle\equiv |0\rangle$ with phonon
vacuum and obtain the excited states by means of the QRPA state
operator, eq. (\ref{eq:StateOp}), $|J_fM_f\rangle\equiv
Q^+_{\lambda \mu i}|0\rangle$, which leads us to the commutator
relation
\begin{equation}
\rho_{if}^T(\vec{r})=\sum_{j_1j_2;\lambda\mu}{\left(i^{\lambda}Y_{\lambda\mu}(\hat{r})\right)^\dag\rho^{\lambda T}_{j_1j_2}(r)} \langle 0|\left[Q_{\lambda \mu
i},\Gamma^+_{\lambda\mu}(j_1j_2)\right]|0\rangle \quad .
\end{equation}
Hence, in QRPA theory the one-phonon transition density is given by
the coherent sum over two-quasiparticle transition densities
entering in the structure of a phonon by the relation:
\begin{equation}
\rho_{\lambda i}^T (r)=\sum_{j_1\geq j_2}{ \rho_{j_1j_2}^{\lambda T}
(r)g^{\lambda i}_{j_1j_2}}\quad . \label{phoph}
\end{equation}

The shape of the transition density defined by eq. (\ref{phoph}) is
rather strongly correlated with the collectivity of the phonon. For
example, the transition densities of the non-collective,
two-quasiparticle excitations typically have pronounced maxima
inside the nucleus. Those corresponding to the collective
transitions with a large number of coherently contributing
two-quasiparticle transitions have a maximum at the nuclear surface.

The reduced transition probability $B(E\lambda)$ for the excitation of a state $J_f$ from the ground state $J_i$ is connected with the transition density with the relation:
\begin{equation}
B(E\lambda)=\frac{2J_f+1}{2J_i+1}\left[\sum^{1}_{T=0}{e^\lambda_T\int_{0}^{\infty}r^{\lambda}\rho_{\lambda i}^T (r)r^{2}dr}\right]^{2},
\end{equation}
where $e^{\lambda}_T$ denotes the effective isoscalar and isovector charges, respectively, introduced before.

\subsection{The QPM Model Parameters}\label{ssec:Params}

Following ref. \cite{Vdo,Bohr} the ratio
$\kappa^{(\lambda)}_{1}$/$\kappa^{(\lambda)}_{0}$ of the isovector
and isoscalar multipole strength parameters, respectively, is
assumed to be a constant, independent of the multipolarity
$\lambda$. We can find this ratio from the dipole coupling constants
by projecting the spurious 1$^{-}$ state to zero excitation energy
and fitting the experimental energy of the Giant Dipole Resonance
(GDR)\cite{Berman:75,Adrich:2005}. For those nuclei, where GDR data
are not available the empirical $E^{GDR}_{max}=76/A^{1/3}$ law is
used. The E1 transition matrix elements are calculated with
recoil-corrected effective charges $q_{n}=-Z/A$ for neutrons and
$q_{p}=N/A$ for protons, respectively, as discussed in sect.
\ref{sec:Features} $^{[3]}$. \footnotetext[3]{Note, that in our
previous work \cite{nadia:2004a} the dipole response in
$^{120-132}$Sn was calculated with the bare charges, leading to
systematically smaller values of the total transition strength.}

\section{Results for the Dipole Response}\label{sec:results}

\subsection{General Features of the Dipole Response}\label{ssec:GenDip}

\begin{figure*}
\includegraphics[width=17cm,angle=0]{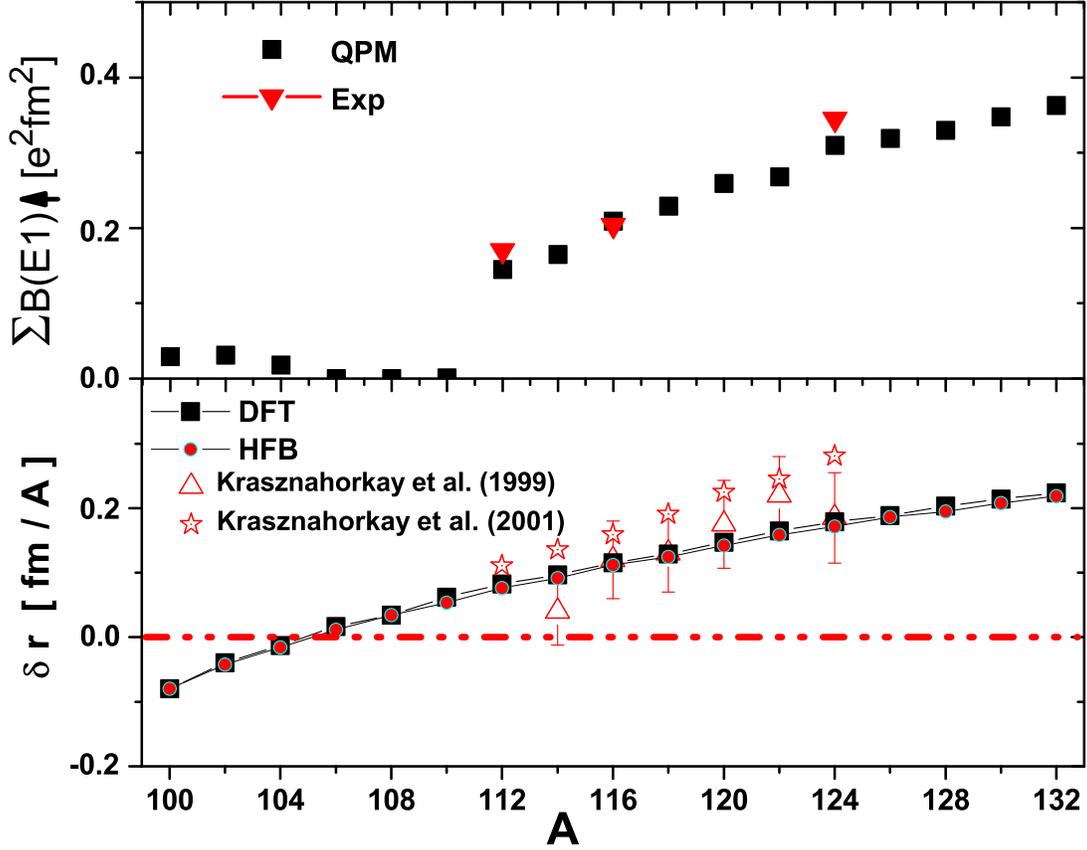}\\
\caption{(Color online) QPM results for the total PDR strengths in the
$^{100-132}$Sn isotopes (upper panel) are displayed for comparison
together with the nuclear skin thickness $\delta r$, eq.
\protect(\ref{eq:skinthick}) (lower panel). Experimental data on the
total PDR strengths in $^{116}$Sn and $^{124}$Sn
ref.\cite{Govaert98} and $^{112}$Sn \cite{Ozel} are also shown. In
the lower panel, the skin thickness derived from charge exchange
reactions by Krasznahorkay et al. \protect\cite{Sn-skin1,Sn-skin2}
are indicated.} \label{FIG:FIG3}
\end{figure*}

The dependence of the calculated total PDR strength ($\sum B(E1)\uparrow$) on the mass
number for the whole chain of isotopes $^{100-132}$Sn is shown in
Fig.\ref{FIG:FIG3} and compared to the skin thickness $\delta r$, eq.
(\ref{eq:skinthick}) of these nuclides. Here the sum is taken over
QRPA dipole states presented in Table 1 and Table 2, respectively.
According to state vectors structure they have been associated with PDR.
These results illustrate and
confirm the conclusion drawn in sect. \ref{sec:Features} and as
already stated in previous work \cite{nadia:2004a} establish the
close relationship of the PDR strength and the skin thickness.

In the region between $^{110-132}$Sn the total PDR strength increases
smoothly with the neutron number. This establishes a clear
correlation of the total PDR strength and the thickness of the
neutron skin in these nuclei, thus confirming our previous results
\cite{nadia:2004a,nadia:2004b,Volz:2006} and the more recent
investigations for several $N=82$ isotones \cite{Volz:2006}. The
close relationship between $\delta r$ and the PDR modes is
underlined by the result, that the PDR becomes negligibly small in
the region $^{106-108}$Sn, where $\delta r$ changes sign
(see Fig.\ref{FIG:FIG3}).
In these isotopes the lowest-lying states carry the characteristics
of the low-energy branch of the GDR as indicated by the structure
and shape of the transition densities.

From the QRPA calculations in $^{110-132}$Sn a sequence of low-lying
one-phonon 1$^{-}$ states at excitation energies $E^*= 6- 7.5$~MeV
of almost pure neutron structure is obtained with a minor fraction
of protons less than 1\%. The structure of the state vectors is
indicated in Table I. The most important part of the
total PDR strength comes from the excitations of the least bound
neutrons from the 3s- and 2p- and 2d-subshells. Some other neutron
orbitals of significance for the size of the neutron skin are
$1h_{11/2}$ and $1i_{13/2}$, which have an important contribution to
the PDR transition matrix elements in $^{132}$Sn nuclei, for
example.

The dominant neutron structure and remarkable stability of the
wave functions of the low-lying one-phonon 1$^{-}$ states in these
nuclei is in agreement with our previous findings on the PDR mode in
$^{120-132}$Sn isotopes \cite{nadia:2004a} and N=82 \cite{Volz:2006}
isotones.

Towards the lighter Sn isotopes the average energy of the excited
dipole states increases, while their total number decreases (see
Table I). The dependence of the PDR energy on the
neutron excess is connected with the one-neutron separation energy,
decreasing  gradually towards the heavier tin isotopes. Such a tendency has
been observed also experimentally in N=82 isotones (see ref.
\cite{Tonchev:2007}).

An exception is the double magic $^{132}$Sn nucleus, in which the PDR
centroid energy is $E^*=7.1 MeV$ and is still below the
neutron emission threshold. The present result in $^{132}$Sn is
slightly different from our previous one
\cite{nadia:2004a,nadia:2004b} due to minor readjustments in the
single-particle spectrum.

In $^{100-104}$Sn the lowest dipole excitations, E$^*$= 8.1-8.3 MeV,
are dominated by proton excitations. The structure of the QRPA state
vectors and B(E1) transition probability are given in Table II.
 There it is seen, that configurations involving
quasibound $2p_{3/2}$ and $1g_{9/2}$ proton states confined by the
Coulomb barrier are the major components. This is a remarkable
Coulomb effect enlightening the delicate balance among various
effects as a prerequisite for a PDR and, by comparison to Fig.
\ref{FIG:FIG2}, the existence of a nuclear skin. From Fig.
\ref{FIG:FIG2} it is seen, that this is the mass region, where the
neutron skin turns into a proton skin. In agreement with the
considerations in sect. \ref{sec:Features} the vanishing skin is
accompanied by a strong suppression of the dipole strength. The
smallest strength is found at $A=110-112$, which is slightly above
the turnover point of $\delta r$ at $A=106$. This delay is caused by
Coulomb effects, which enhance the dipole response from weakly bound
proton orbitals in that mass region over the values to be expected
for full isospin symmetry.

Electromagnetic breaking of isospin symmetry is also the main reason
for the persisting of low-energy dipole strength close to
$^{100}$Sn. Already the quite different behavior is an indication
for another mechanism underlying these excitations. There, at $N=Z$
the isoscalar dipole charge vanishes, hence the electromagnetic
operator by itself does no longer support isoscalar transition.
However, Coulomb effects in the single-particle wave functions
translate into an intrinsic isospin symmetry breaking on the level
of matrix elements. The mechanism behind a neutron skins in the
heavy Sn isotopes is a strong interaction effect, namely the
repulsive action of the isovector self-energy to the neutron
mean-field. In neutron-rich nuclei the isovector self-energy adds
attractively to the proton potential, which partially compensates
the Coulomb repulsion. Because for N$\to$Z the isovector self-energy
becomes negligible the proton skins seen in the light Sn isotopes
must be of a different origin. In fact, they are due to the Coulomb
potential. Towards $N=Z$ the Coulomb interaction can act in full
strength on the protons, pushing them apart and leading to a
rearrangement of a certain fraction of the nuclear charge in the
surface region.

The average energies in Table I and Table II
have been obtained by the relation $\left\langle
E\right\rangle=\sum_{i} E_{i}B_{i}/\sum_{i} B_{i}$, where $E_{i}$
and $B_{i}$ are the QRPA energies and reduced transition
probabilities, respectively.

\subsection{PDR and GDR Transition Densities}\label{ssec:DipTRD}

\begin{figure*}
\includegraphics[width=16cm,angle=0]{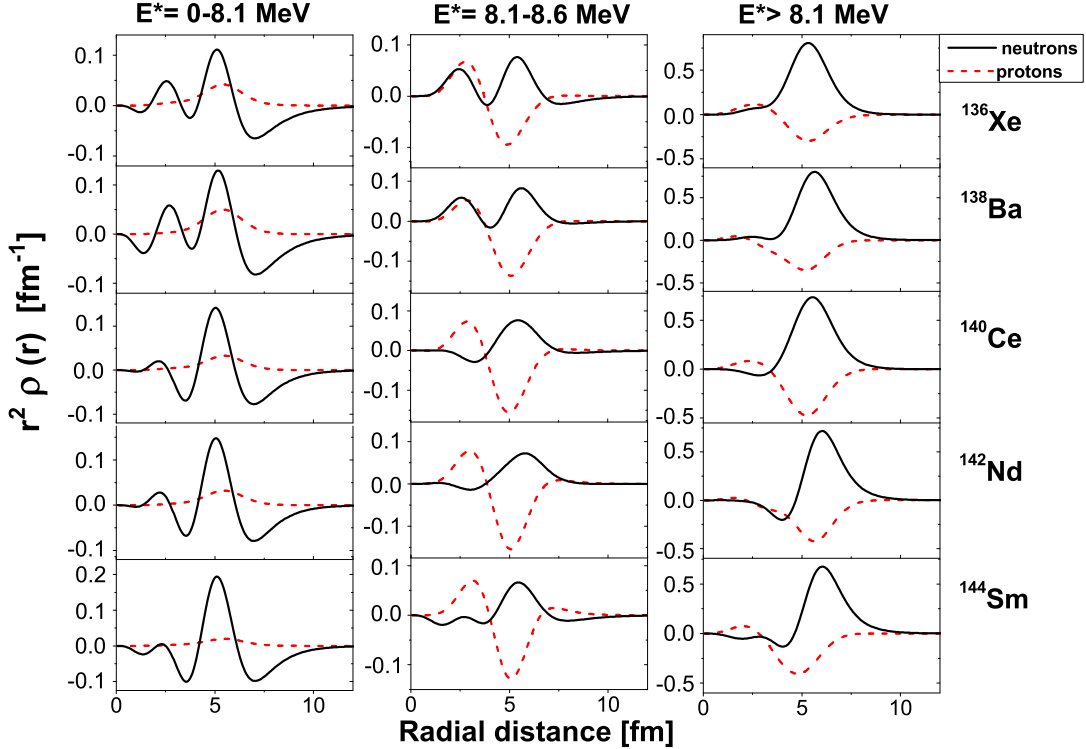}\\
\caption{(Color online) QRPA results for the one-phonon dipole transition densities
in N=82 nuclei. } \label{FIG:FIG4}
\end{figure*}

\begin{figure*}
\includegraphics[width=14cm,angle=0]{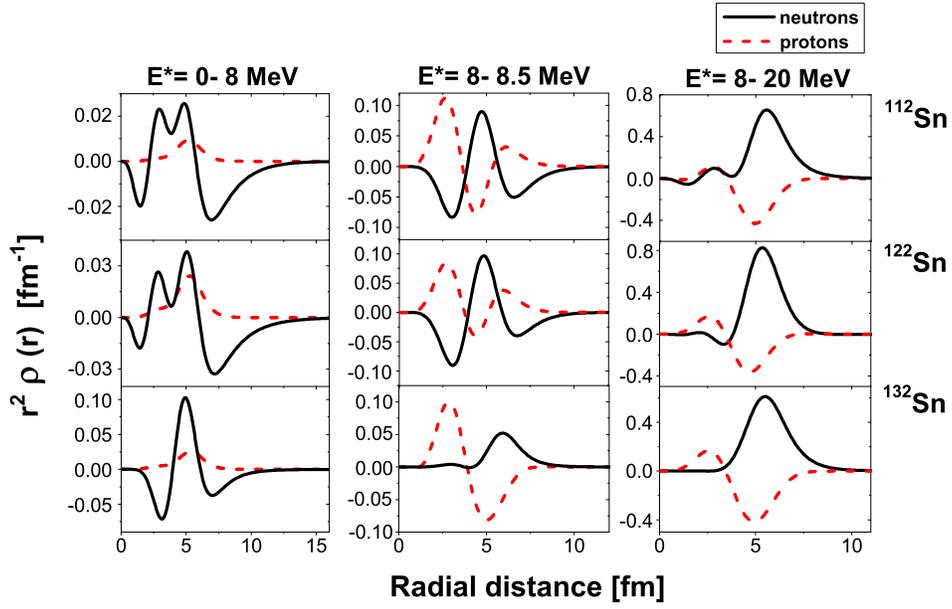}\\
\caption{(Color online) QRPA results for the one-phonon dipole transition densities
in $^{112,122,132}$Sn nuclei. } \label{FIG:FIG5}
\end{figure*}

\begin{figure*}
\includegraphics[width=13cm,angle=0]{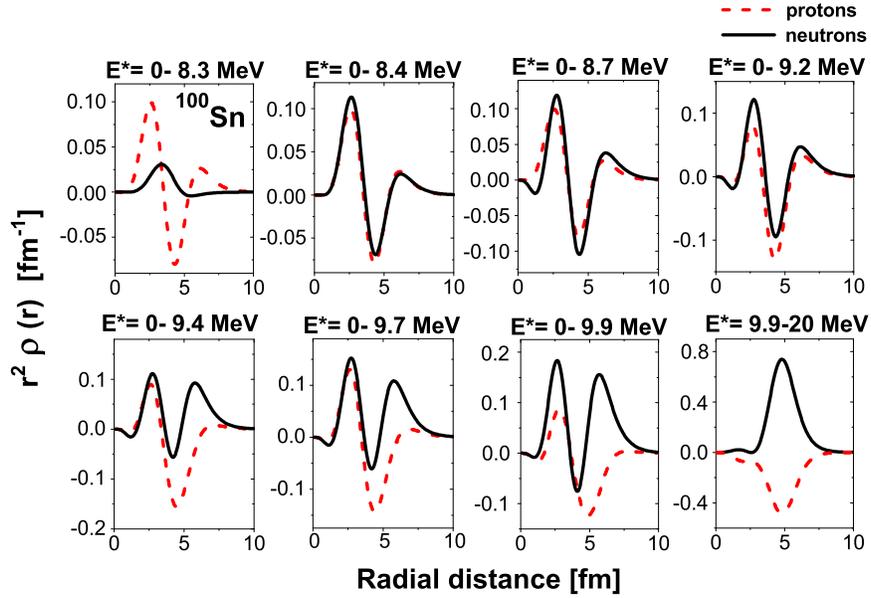}\\
\caption{(Color online) QRPA results for the one-phonon dipole transition densities
in $^{100}$Sn. } \label{FIG:FIG6}
\end{figure*}

For a more detailed insight into the characteristic features of the
dipole excitations we consider the evolution of the proton and
neutron transition densities for in the various energy regions. In
Fig.\ref{FIG:FIG4} and Fig.\ref{FIG:FIG5} we display the QRPA transition
densities for several $N=82$ isotones in and the $Z=50$ isotopes
$^{112,122,132}$Sn for three different regions of excitation
energies: the low-energy PDR region below the neutron emission
threshold, the transitional region up to the GDR and in the GDR
region and beyond$^{4}$. \footnotetext[4]{For a detailed discussion of the dipole response of the N=82 isotones we refer to ref. \cite{Volz:2006}.} The
transition densities displayed in Fig.\ref{FIG:FIG4}, Fig. \ref{FIG:FIG5} and Fig. \ref{FIG:FIG6} were obtained by
summing over the transition densities of the individual one-phonon states located in the energy intervals denoted at the top of each column of the figures, i.e.  
\begin{equation}
\rho_{\lambda }^T (r)=\sum_{i}\rho_{\lambda i}^T(r).
\end{equation}
$\rho_{\lambda i}^T (r)$ is determined by equation (\ref{phoph}), where the module and the phases are unambiguously determined by our microscopic approach. The neutron and proton transition densities are then obtained by taking half the difference and the sum of the isoscalar and isovector pieces, respectively.

A common features of the all cases presented in Fig.4 is that up to $E^*= 8.1$~MeV the
protons and neutrons oscillate in phase in the nuclear interior,
while at the surface only neutron transitions contribute. The same behavior of the neutron and proton transition densities is observed below 8 MeV for tin isotopes (see Fig.5). This
pattern is generic to the lowest dipole states making it meaningful
to distinguish these excitations from the well known GDR states.
Hence, we are allowed to identify the PDR states with a new mode of
nuclear excitation, not seen in stable $N\sim Z$ nuclei.

In the energy region $E^*=8.1-8.6$~MeV for N=82 isotones and 8-8.5 MeV for Z=50 respectively,
the transition densities suddenly
change. Rather abruptly, protons and neutrons start to oscillate out
of phase over the whole nuclear volume as known from the GDR. Thus,
we are encountering the low-energy part of the GDR, although the
strengths of these two different type of excitations, the PDR and
the low-energy GDR tail are quite comparable. Also energetically
they are located very close to each other. This makes the task to
distinguish the two modes rather demanding. Theoretically, we can
always use the transition densities for a detailed analysis and a
precise identification of the mode although a corresponding
experimental measurement will not be feasible for the foreseeable
future. Finally, we show in Fig.\ref{FIG:FIG4} and Fig.\ref{FIG:FIG5}
the one-phonon QRPA proton and neutron transition densities for the states at
$E^*>8.1$~MeV and $E^*>8$~MeV, respectively. A pronounced isovector oscillation
of protons against
neutrons, peculiar to the GDR, is observed. The latter, as a very
collective mode, has a strength one order of magnitude larger than
the PDR.

We have pointed out the special character of the low-energy dipole
excitations in the $^{100-104}$Sn isotopes. This is reflected also
in the transition densities. In $^{100}$Sn, Fig.\ref{FIG:FIG6}, proton
oscillations prevail below $E^*<8.4$ MeV. In the nuclear interior
isoscalar, or mixed symmetry vibrations of protons and neutrons are
found, while at the surface only protons contribute. Hence, this
mode could be related to a proton skin excitations. In the energy
region $E^*=8.4-9.2$ MeV oscillations of weakly bound neutrons from
the surface region take place. The behaviour of the transition
densities and the structure of the $1^{-}$ states at these energies
is similar to the neutron PDR mode identified in the more neutron
rich tin isotopes (see Fig.\ref{FIG:FIG5} as well). At energies
$E^*>9.2$ MeV the low-energy tail of the GDR is encountered. The
last plot in Fig.\ref{FIG:FIG6} displays the neutron and proton
transition densities summed over the GDR region, $E^*=9.2- 20$ MeV.

\begin{center}
\begin{figure*}
\includegraphics[width=18cm,angle=0]{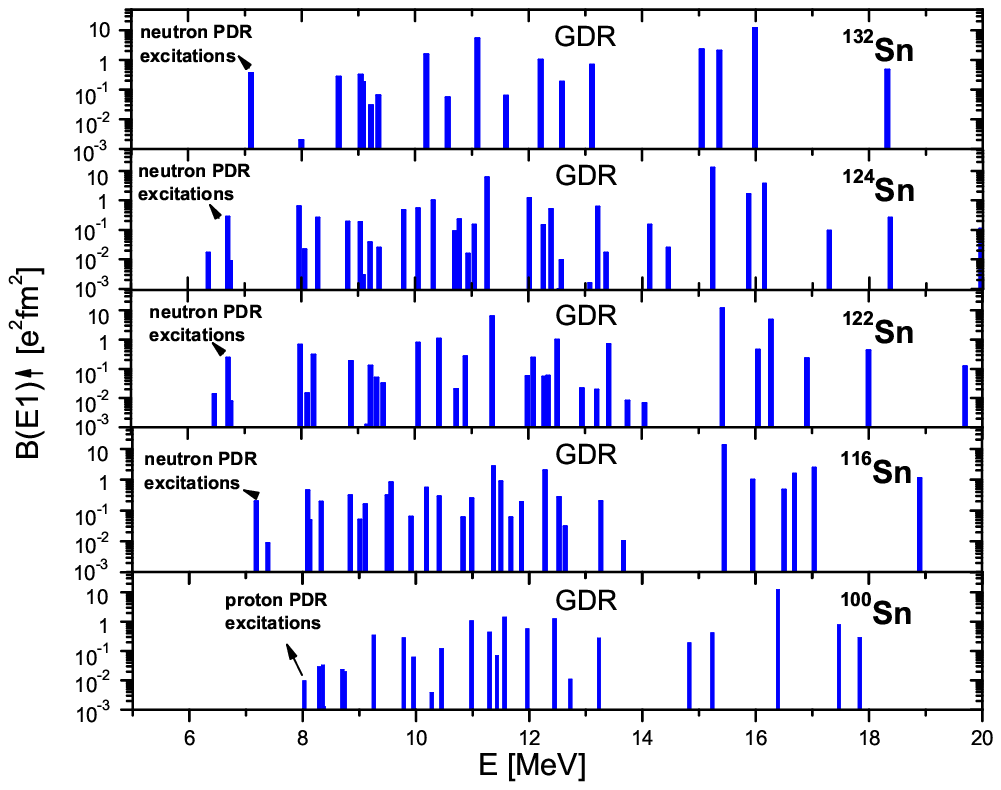}\\
\caption{(Color online) QRPA results for the PDR and GDR strength distributions in
$^{100,116,122,124,132}$Sn isotopes.} \label{FIG:FIG7}
\end{figure*}
\end{center}

The QRPA calculations on the  neutron and proton PDR and the GDR strength distributions at excitation energies up to 20 MeV in several thin isotopes in the mass region $^{100}$Sn$\div^{132}$Sn are presented in Fig.(\ref{FIG:FIG7}).
An interesting feature we have observed is that between the N=50 and N=82 closed shells with the increase of the neutron number from $^{100}$Sn toward $^{132}$Sn the PDR strength is shifted to lower excitation energies relatively to the GDR mode, which is almost unchanged. This can be explained with a strong correlation between the PDR excitation energy and the energy of the neutron threshold, which also decreases in the same direction.  
\begin{table*}\label{tab:tab1}
\caption{Energy, B(E1) values and wave functions of the first QRPA
1$^{-}$ states in the $^{110\div132}$Sn isotopes. Only the dominant
neutron and proton components are given. Neutron and proton
configurations are denoted by the indices $\nu$ and $\pi$,
respectively. }
\begin{tabular}{|c|c|c|c|c|cc|}
\hline\hline
Nucleus & State & Energy &
Structure & B(E1)$\uparrow$& $<E>$& \\
&J$_\nu ^\pi $  & [MeV] & $w_{j_1j_2}$, $\%$ & [e$^{2}$fm$^{2}$]&[MeV] &\\ \hline
&  &  &  &  & & \\
$^{110}$Sn & 1$_1^{-}$ & 7.834 & 99.9$\%[1g_{7/2}2f_{7/2}]\nu$ &
0.001 & 7.8& \\
\hline
$^{112}$Sn & 1$_1^{-}$ & 7.509 &
99.8$\%[1g_{7/2}2f_{7/2}]\nu$
& 0.001 & &\\
& 1$_2^{-}$ & 7.906 & 99.0$\%\left[3s_{1/2}3p_{3/2}\right]\nu$ & 0.144
& 7.9& \\
\hline
$^{114}$Sn & 1$_1^{-}$ & 7.329 & 99.8$\%[1g_{7/2}2f_{7/2}]\nu$ & 0.001 & &\\
& 1$_2^{-}$ & 7.665 & 99.2$\%\left[3s_{1/2}3p_{3/2}\right]\nu$ & 0.159
& 7.7& \\
& 1$_3^{-}$ & 8.021 & 99.9$\%\left[2d_{3/2}3p_{3/2}\right]\nu$ & 0.005 & &\\
\hline
$^{116}$Sn & 1$_1^{-}$ & 6.974 & 99.7$\%[2g_{7/2}3f_{7/2}]\nu$ & 0.001 & &\\
& 1$_2^{-}$ & 7.188 & 99.$\%\left[3s_{1/2}3p_{3/2}\right]\nu$ & 0.199
& 7.2&\\
&  &  & + 0.1$\%\left[1g_{9/2}1h_{11/2}\right]\pi$ & & &\\
& 1$_3^{-}$ & 7.391 & 99.9$\%\left[2d_{3/2}3p_{1/2}\right]\nu$ & 0.009 & &\\
\hline
$^{118}$Sn & 1$_1^{-}$ & 6.904 & 99.6$\%[1g_{7/2}2f_{7/2}]\nu$ & 0.001 & &\\
& 1$_2^{-}$ & 7.054 & 98.1$\%\left[3s_{1/2}3p_{3/2}\right]\nu$ & 0.208
& 7.1& \\
&  &  & + 0.1$\%\left[1g_{9/2}1h_{11/2}\right]\pi$ & & & \\
& 1$_3^{-}$ & 7.098 & 99.1$\%\left[2d_{3/2}3p_{3/2}\right]\nu$ & 0.02 & &\\
\hline
$^{120}$Sn & 1$_1^{-}$ & 6.795 & 99.6$\%[2d_{3/2}3p_{3/2}]\nu$ & 0.009 & &\\
& 1$_2^{-}$ & 6.870 & 95.2$\%\left[1g_{7/2}2f_{7/2}\right]\nu$ & 0.014
& 6.9& \\
& 1$_3^{-}$ & 6.910 & 94$\%\left[3s_{1/2}3p_{3/2}\right]\nu$ & 0.238 & &\\
&  &  & + 0.1$\%\left[1g_{9/2}1h_{11/2}\right]\pi$ & & &\\
\hline
$^{122}$Sn & 1$_1^{-}$ & 6.469 & 99.8$\%[2d_{3/2}3p_{3/2}]\nu$ & 0.014 & &\\
& 1$_2^{-}$ & 6.710 & 95.3$\%\left[3s_{1/2}3p_{3/2}\right]\nu$ & 0.245
& 6.7& \\
&  &  & +  0.1$\%\left[1g_{9/2}1h_{11/2}\right]\pi$  & & &\\
& 1$_3^{-}$ & 6.754 & 95.8$\%\left[1g_{7/2}2f_{7/2}\right]\nu$ & 0.009 & &\\
&  &  & +  0.1$\%\left[1g_{9/2}1h_{11/2}\right]\pi$ & & &\\
\hline
\hline
\end{tabular}
\end{table*}
\begin{table*}
\begin{tabular}{|c|c|c|c|c|cc|}
& & & Table I continued & & &\\
\hline\hline
$^{124}$Sn & 1$_1^{-}$ & 6.359 & 99.8$\%[2d_{3/2}3p_{3/2}]\nu$ & 0.017 & &\\
& 1$_2^{-}$ & 6.702 & 94.8$\%\left[3s_{1/2}3p_{3/2}\right]\nu$ & 0.284
 & 6.68 &\\
&  &  & + 0.1$\%\left[1g_{9/2}1h_{11/2}\right]\pi$ & & & \\
& 1$_3^{-}$ & 6.749 & 95.6$\%\left[1g_{7/2}2f_{7/2}\right]\nu$ & 0.009 & & \\
&  &  & + 0.1$\%\left[1g_{9/2}1h_{11/2}\right]\nu$ & & & \\
\hline
$^{126}$Sn & 1$_1^{-}$ & 6.180 & 99.7$\%[2d_{3/2}3p_{3/2}]\nu$ & 0.019 & &\\
& 1$_2^{-}$ & 6.621 & 51.4$\%\left[3s_{1/2}3p_{3/2}\right]\nu$ &
0.163 & 6.6 &\\
&  &  & + 48.3$\%\left[1g_{7/2}2f_{7/2}\right]\nu$ & & &\\
& 1$_3^{-}$ & 6.642 & 51.5$\%\left[1g_{7/2}2f_{7/2}\right]\nu$ & 0.137 & & \\
&  &  & + 47.$\%\left[3s_{1/2}3p_{3/2}\right]\nu$  & & &\\
&  &  & + 0.2$\%\left[1g_{9/2}1h_{11/2}\right]\pi$  & & &\\
\hline
$^{128}$Sn & 1$_1^{-}$ & 5.611 & 99.7$\%[2d_{3/2}3p_{3/2}]\nu$ & 0.023 & & \\
& 1$_2^{-}$ & 6.201 & 97.8$\%\left[3s_{1/2}3p_{3/2}\right]\nu$ & 0.306
& 6.2 &\\
&  &  & + 0.2$\%\left[1g_{9/2}1h_{11/2}\right]\pi$ & & & \\
& 1$_3^{-}$ & 6.352 & 99.1$\%\left[1g_{7/2}2f_{7/2}\right]\nu$ & 0.001 & &\\
\hline
$^{130}$Sn  & 1$_1^{-}$ & 5.172 & 99.7$\%[2d_{3/2}3p_{3/2}]\nu$ & 0.028 & & \\
& 1$_2^{-}$ &5.882 & 98.1$\%\left[3s_{1/2}3p_{3/2}\right]\nu$ & 0.319
& 5.8 &\\
&  &  & +
0.2$\%\left[1g_{9/2}1h_{11/2}\right]\pi$ & & & \\
& 1$_3^{-}$ & 6.114 & 99.4$\%\left[1g_{7/2}2f_{7/2}\right]\nu$ & 0.0002 & &\\
\hline
$^{132}$Sn  & 1$_1^{-}$ & 5.754 & 99.7$\%[1g_{7/2}2f_{7/2}]\nu$ &  0.0001 & &\\
& 1$_2^{-}$ & 7.109 & 88.6$\%\left[2d_{5/2}2f_{7/2}\right]\nu$ & 0.363
& 7.1 &\\
&  &  & + 10.8$\%\left[1h_{11/2}1i_{13/2}\right]\nu$ & & & \\
&  &  & + 0.2$\%\left[1g_{9/2}1h_{11/2}\right]\pi$ & & & \\
 \hline\hline
\end{tabular}
\end{table*}

\begin{table*}\label{tab:tab2}
\caption{The same as Table I for $^{100\div104}$Sn isotopes.}
\begin{tabular}{|c|c|c|c|c|cc|}
\hline\hline
Nucleus & State & Energy &
Structure & B(E1)$\uparrow$&  &$<E>$ \\
&J$_\nu ^\pi $  & [MeV] & $w_{j_1j_2}$,$\%$ & [e$^{2}$fm$^{2}$]&  &[MeV] \\ \hline
&  &  &  &  &  & \\
$^{100}$Sn & 1$_1^{-}$ & 8.032& 99.5$\%[1f_{5/2}2d_{5/2}]\nu$
& 0.001 &  &\\
& 1$_2^{-}$ & 8.292 & 82.1$\%\left[2p_{3/2}2d_{5/2}\right]\pi$ & 0.028
& &8.29 \\
\hline
$^{102}$Sn & 1$_1^{-}$ & 8.174 &
82.6$\%[2p_{3/2}2d_{5/2}]\pi$ & 0.031
& & \\
\hline
$^{104}$Sn & 1$_1^{-}$ & 8.256 &
80.8$\%[2p_{3/2}2d_{5/2}]\pi$ & 0.016
& & \\
\hline\hline
\end{tabular}
\end{table*}

\subsection{Multi-phonon effects in the low-energy dipole spectra of $^{124}$Sn.}

In the multi-phonon QPM calculations the structure of the excited states is described by wave functions as defined in eq. \ref{wf}.We now investigate multi-phonon effects using a model space with up to three-phonon components, built from a basis of QRPA states with J$^\pi=1^{\pm}, 2^+, 3^-, 4^+, 5^-, 6^+, 7^-, 8^+$. Since the
one-phonon configurations up to $E^*$=20~MeV are considered the
core polarization contributions to the transitions of
the low-lying 1$^{-}$ states are taken into account explicitly.
Hence, we do not need to introduce dynamical effective charges.
In the excitation energy interval up to $E^*$=9 MeV we use a total of about 250 multi-phonon configurations.
\begin{center}
\begin{figure*}
\includegraphics[width=17cm,angle=0]{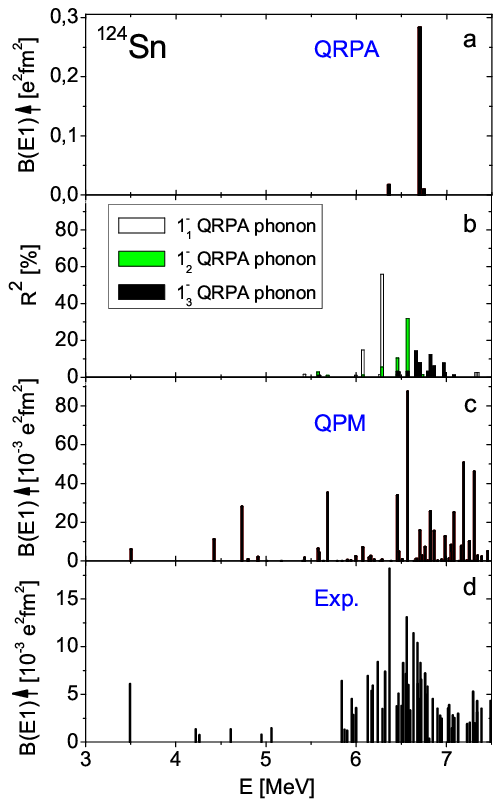}\\
\caption{(Color online) Electromagnetic dipole response in $^{124}$Sn from: a/QRPA calculations; c/QPM with up to three-phonon configuration space including 250 components; d/experimental data from \cite{Govaert98} up to excitation energies $E^*$=7.5 MeV. 
Distributions of the PDR QRPA phonons over the 1$^-$ excited states is presented in Fig 10.b.}
\label{FIG:FIG8}
\end{figure*}
\end{center}
The results for the dipole response below the neutron threshold in $^{124}$Sn are presented in Fig.\ref{FIG:FIG8}. By comparing Fig.\ref{FIG:FIG8}a and Fig.\ref{FIG:FIG8}c it is seen that the pure two-quasiparticle QRPA strengths is strongly fragmented once the coupling to multi-phonon configurations is allowed. As found previously the lowest-lying 1$^{-}$ state without a QRPA counterpart is predominantly given by a two-phonon quadrupole-octupole excitation \cite{PDRrev:06}. The $[2^{+}_{1}\otimes3^{-}_{1}]$ configuration accounts for 85\% of the QPM wave function. The $1^-_1$ state is located at $E^*=3.50$~MeV, carrying a reduced transition probability $B(E1;g.s \rightarrow 1^{-}_{1})=6.06$ $10^{-3}$ $e^{2}fm^{2}$. The values are in a good agreement with the experiment ($E^*=3.49$~MeV and $B(E1;g.s \rightarrow 1^{-}_{1})=6.08$ $10^{-3}$ $e^{2}fm^{2}$ \cite{Bry} and previous QPM calculations refs.\cite{Bry,nadia:2004a}.

Here, our attention is especially focused on the 1$^{-}$ states above the
two-phonon dipole state and below the neutron threshold. From the analysis of the QRPA calculations discussed above the 1$^{-}$ states presented in Fig.\ref{FIG:FIG8}a are PDR modes.
Their fragmentation over the multi-phonon 1$^{-}$ excited states are shown in Fig.\ref{FIG:FIG8}b.
From that plot we can determine the energy region, where the PDR is located.
In the particular case of $^{124}$Sn in the interval $E^* \leq 7.5$~MeV is exhausted about 80$\%$ of the total one-phonon PDR strength.

Comparing the fragmentation pattern of the theoretical low-energy dipole strength in $^{124}$Sn to recent measurements \cite{Bry,Govaert98}, displayed in Fig.\ref{FIG:FIG8}d, we find that the three-phonon QPM results are still not fully accounting for the observed distribution, although we have increased the number of the multi-phonon configurations twice in comparison with our previous calculations from \cite{nadia:2004a}. However, the calculated total QPM dipole strength in the PDR energy range $E^*= 5.7-7.2$~MeV is
$\sum B(E1)_{QPM}=0.324$ $e^{2}fm^{2}$ which is almost identical to the experimentally deduced strength, $\sum B_{exp}(E1)=0.345(43)e^{2}fm^{2}$.

The good overall agreement between the calculations and the experiment \cite{Govaert98} for the total PDR strength and centroid energy in $^{124}$Sn indicates that in this particular case the main PDR properties could already be determined on the level of the one-phonon approximation. A similar conclusion was drawn in our previous QPM calculations for several Sn isotopes with A=120-130 \cite{nadia:2004a}. An important observation is that the low-energy  tail of the GDR can give a strong contribution to the dipole strength around the particle threshold. This effect appears because the GDR states may overlap with the PDR region and can be fragmented due to coupling to multi-phonon states. The effect becomes increasingly important in nuclei where the neutron threshold is higher, hence approaching the GDR region, as in the lighter Sn isotopes or $^{132}$Sn, where the PDR strength is situated very close to the neutron threshold. Such a situation demands much larger model spaces.

\section{Other Model Calculations and Experimental Data}\label{sec:OthersData}
\subsection{PDR Models}
Overall, the present Sn results are in, at least qualitative,
agreement with the theoretical PDR investigations by Density
Functional Theory (DFT) \cite{Cha94}, relativistic RPA \cite{Piek},
relativistic QRPA \cite{Vret01,Paar}, extended theory of finite
Fermi systems \cite{THartmann} and QRPA-PC (Quasiparticle Random Phase Approximation plus Phonon Coupling)\cite{Sarchi} . They confirm the conclusions drawn
from our former QPM calculations in $^{208}$Pb \cite{Rye02} and for
the N=82 case, studied recently in \cite{Volz:2006}. All these
different approaches confirm the PDR mode as a universal low-energy
dipole mode of a character generic for isospin asymmetric nuclei.

As an example we cite the studies of ref.\cite{Cha94} investigating
low-lying dipole states in $^{40-48}$Ca in a density functional
theory approach. Similar to the much heavier nuclei considered here
the PDR is predicted to be located in the energy range 5-10 MeV.
Also in that nucleus, the centroid energy of the PDR strength is
found to decrease with the number of the neutrons, while the
integrated PDR strength (below the neutron particle emission
threshold) increases. These results agree with the present calculations and our findings
in refs. \cite{nadia:2004a, nadia:2004b, Volz:2006} in the Sn
isotopes.
A common result of all model calculations discussed here is a clear
connection between the existence of low-energy dipole strength and the presence of isospin assymetry
or nuclear skin in the investigated nuclei. However, we emphasize, that arguments
based on the energy alone are likely to be insufficient for a
unambiguous identification of the dipole states as belonging to the
PDR. To our understanding, as an important conclusion from the
analysis of the transition densities, the PDR strength is attached
only to the states located below the neutron particle emission
threshold. Hence, the centroid of the PDR energy has a tendency to
be closely connected with the one-neutron separation energy. At
higher energies, the dipole spectrum merges rapidly into the low
energy tail of the GDR and the transitions lose their characteristic
PDR features.

A controversial question is the degree of the collectivity of the
PDR transitions. This issue has been discussed by several authors
\cite{Cha94,Piek,Vret01,Sarchi,Paar07}. In the non-relativistic models like ours \cite{Rye02,nadia:2004b,Volz:2006} and QRPA-PC for example, the PDR is referred to as the excitation of two-quasiparticles states. In $^{132}$Sn the
relativistic QRPA \cite{Vret01} predicts a collective neutron state
at $E^*=8.6$~MeV that has been related to the PDR excitation.
This state contains particle-hole configurations
accounting for transitions into continuum states. The collectivity
of such excitations we found to dependent strongly on the choice of
the spin-orbit potential affecting the energy gap between the bound
hole and unbound particle region by shifting the continuum states.
Our standard choice for the spin-orbit potential strength \cite{Tak}, otherwise describing the spectra reasonably well, disfavors such admixtures. 

Experimental data for low-energy dipole states below the particle
emission threshold are available for a number of Sn isotopes,
$^{116}$Sn and $^{124}$Sn \cite{Govaert98} and recent measurements
in $^{112}$Sn \cite{Ozel}, and for several N=82 isotones
\cite{Volz:2006}. Altogether, our calculations describe these data
quite satisfactory.

\subsection{Dipole Response in $^{130,132}$Sn}\label{ssec:132Sn}

We pay special attention to the region around $^{132}$Sn, because of
the expected closure of the N=82 neutron shell as indicated e.g. by
the energy of the first $2^+$ state. The HFB calculations predict a
double shell closure for protons and neutrons, respectively. On the
other hand, the HFB calculations show, that the N=82 neutron shell
closure depends to some extent on the balance between spin-orbit
splitting and the effective pairing strength.

The LAND-FRS collaboration at GSI has recently measured in a
pioneering Coulomb dissociation experiment the dipole response above
neutron threshold in $^{130,132}$Sn \cite{Adrich:2005}. These
measurements are providing the first data on the dipole response in
the these exotic nuclei. However, we have to be aware, that any
dipole strength below the particle emission threshold -- if existing
-- cannot be accessed by this type of measurement. Besides the GDR a
prominent feature of the data is a resonance-like structure around
$E^*\sim$10 MeV exhausting a few percent of the EWSR in
$^{130,132}$Sn nuclei. In \cite{Adrich:2005} this part of the
response function was interpreted as a PDR. We compare our
calculated integrated dipole photoabsorption cross sections $\sigma$
in the Sn isotopes to the LAND-FRS data \cite{Adrich:2005}) in Table
\ref{tab:tab3}.

A different conclusion is obtained by analyzing our QRPA wave
functions and the dipole transition densities. In $^{130,132}$Sn, we
find dipole excitations, carrying the characteristic features of PDR
transitions, below the neutron particle emission threshold, as
indicated in the first three columns of Table \ref{tab:tab3} and in
Fig.\ref{FIG:FIG7}. In the energy domain $E^*=8-12$~MeV, assigned in
\cite{Adrich:2005} as PDR region, we obtain in both nuclei another
concentration of E1 strength (see also Fig. \ref{FIG:FIG7}). However,
because the transition densities show the GDR-type behavior, we
consider this part of the dipole response as the low-energy tail
({\em LET}) of the GDR. Although the LET evolves in close relation
to the neutron excess, but it does not seem to be related to
excitations of the neutron skin.

In fact, there is a simple proportionality between the dipole
photoabsorption cross section, integrated over an energy interval
around $E^*$, and reduced transition strength,
$\int{dE\sigma_\gamma}\sim E^*B(E1,E^*)$, up to a numerical factor
\cite{Bohr}. Exploiting this relation, we have calculated the
integrated dipole photoabsorption cross sections in the LET regions
of $^{130,132}$Sn. In Tab.\ref{tab:tab3} it is seen that the
theoretical results agree rather well with those determined
experimentally in \cite{Adrich:2005}. Within the experimental error
bars, also the full strengths, including excitations up to 20~MeV,
are reasonably well described.

\begin{table*}
\caption{\label{tab:tab3}Dipole response in $^{130,132}$Sn.
Calculated energies and integrated cross sections (columns denoted
by {\em QPM}) in one-phonon approximation are compared with recent measurements (columns denoted
by {\em Exp.}) \protect\cite{Adrich:2005} of PDR and GDR in Sn
isotopes. The calculated integrated PDR and low-energy GDR cross sections are
denoted by $\int\sigma^{PDR}$ and $\int\sigma_{LET}^{GDR}$, respectively. The total photoabsorption cross section up to 20 MeV is denoted by $\int\sigma^{GDR}$}.
\begin{tabular}{ccccccccccccc}
\hline\hline
Nucl.& PDR & $\left\langle E \right\rangle_{PDR}$& $\int\sigma^{PDR}$&$E_{max}^{PDR}$& $\int\sigma^{PDR}$& $E^{GDR}_{LET}$&$\int\sigma^{GDR}_{LET}$ & $E_{GDR}^{max}$ &$E_{GDR}^{max}$& $\int\sigma^{GDR}$ & $\int\sigma^{GDR}$ &\\
 & $(^{Energy}_{region})$ & [MeV]& [mb MeV]& [MeV]& [mb MeV]&[MeV]& {[mb MeV]}& [MeV]&[MeV] &[mb MeV]\\
 & QPM & QPM &QPM &Exp. &Exp. &QPM &QPM &Exp.&QPM&Exp.&QPM  \\
\hline
$^{130}$Sn&0-7.4 &5.8& 8.2& 10.1(7)& 130(55)&8-11 &137.3 &15.9(5) &16. & 1930(300)*&1616\\
$^{132}$Sn&0-8&7.1&10.4&9.8(7)&75(57)&8-11&97.6 &16.1(7)&16.1&1670(420)*&1518\\
\hline\hline
\end{tabular}
\footnotetext{*Data of ref.\cite{Adrich:2005} integrated up to 20 MeV \cite{Adam:2007}. }   \end{table*}

\begin{center}
\begin{figure*}
\includegraphics[width=17cm,angle=0]{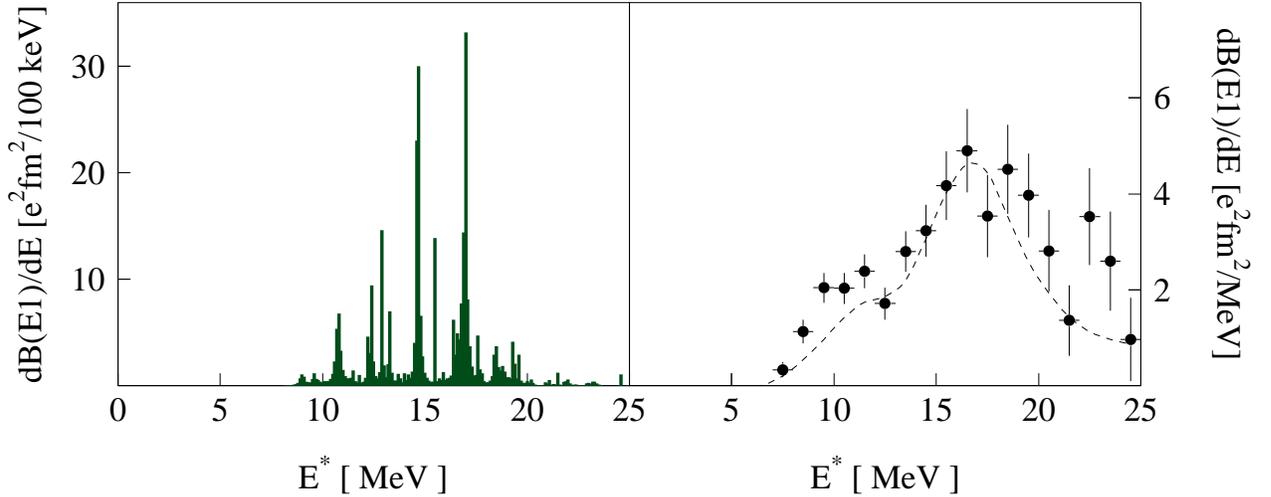}\\
\caption{Electromagnetic QRPA dipole response function and LAND-FRS
data \protect\cite{Adrich:2005} for $^{130}$Sn. The QRPA results
(left), where obtained by solving the Dyson equation and include the
decay widths from particle emission. In the right panel the
theoretical response function has been folded with the experimental
acceptance filter \protect\cite{Adam:2007} (dashed line) and is
compared to the data (symbols).} \label{FIG:FIG9}
\end{figure*}
\end{center}

For the purpose of a realistic description of the measured spectra,
we have applied a slightly different numerical method by solving the
QRPA Dyson equation similar to the approach used in
\cite{PhysRep:97} allowing to take into account explicitly the
continuum decay width $\Gamma^\uparrow$ of the states above particle
threshold, ranging from a few keV up to about hundred keV. Still, a
comparison to the LAND-FRS spectra is only possible after folding
the theory with the experimental acceptance filters
\cite{Adam:2007}. The results of such calculations is shown in Fig.
\ref{FIG:FIG9} and Fig.\ref{FIG:FIG8} with a quite remarkable agreement to the data.

\begin{center}
\begin{figure*}
\includegraphics[width=17cm,angle=0]{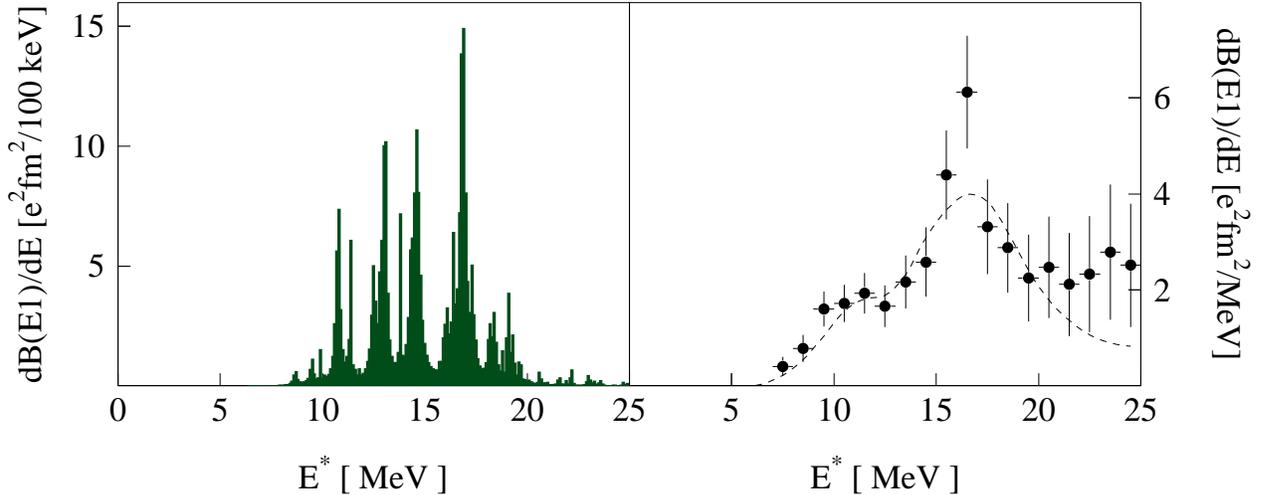}\\
\caption{Electromagnetic QRPA dipole response function and LAND-FRS
data \protect\cite{Adrich:2005} for $^{132}$Sn. As in Fig.
\ref{FIG:FIG9} the QRPA results (left) include the decay widths from
particle emission. In the right panel the theoretical response
function has been folded with the experimental acceptance filter
\protect\cite{Adam:2007} (dashed line) and is compared to the data
(symbols).} \label{FIG:FIG10}
\end{figure*}
\end{center}

\section{Summary and Conclusions}\label{sec:summary}

Low-energy dipole excitations in $^{100-132}Sn$ isotopes were
studied by a theoretical approach based on HFB and QPM theory. From
our calculations in $^{112}$Sn $\div$ $^{132}$Sn we obtained
low-energy dipole strength in the energy region below 8 MeV, close
but below the neutron emission threshold. These states are of a
special character. Their structure is dominated by neutron
components and their transition strength is directly related to the
presence of a neutron skin. Their generic character is further
confirmed by the shape and structure of the related transition
densities, showing that these {\it pygmy dipole resonances} (PDR)
are clearly distinguishable from the conventional GDR mode. Our
calculations show a rather abrupt transition from PDR- to the
GDR-type excitations, typically occurring at energies slightly above
the particle threshold. An important finding in our calculations is that an accurate description of the PDR part of the dipole spectrum requires a single particle spectrum corresponding to a total effective mass $\frac{m^*}{m}=1$. From that observation we conclude that non-localities and dynamical effects form core polarizations are important for a proper description of the PDR spectrum. Pure HF and HFB models, whether non-relativistic or relativistic, typically use effective masses considerably less then unity. Hence, such approaches might miss important effects.

In the most proton-rich exotic nuclei $^{100-104}$ Sn the lowest dipole states
are almost pure proton excitations. They are related to oscillations
of weakly bound protons, indicating a proton PDR. The interesting
point is, that we found these states in heavy nuclei with N slightly
larger, or equal to Z. We suggest, that the effect is due to Coulomb
repulsion, that pushes out weakly bound protons orbitals to the
nuclear surface. Coulomb effects also induce a considerable amount
of isospin breaking at the level of single-particle wave functions.

Since similar observation have been made in the nearby $N=82$
isotonic nuclei, we may conclude, that the features discussed here are
indicating a new universal mode of excitation. It is worthwhile to
extent the investigations also in other mass regions. Promising
candidates are the Ni and Ca isotopes, but also the light mass region,
where a mixing between halo and skin degrees of freedom can be
expected, which may lead to still other modes of excitations.

\subsection*{{\bf Acknowledgements}} This work was supported by DFG
contract Le-439/6 and GSI. We are grateful to Adam Klimkiewicz for providing us with experimental results of the photoabsorption cross sections in $^{130,132}$Sn and
his support in performing the folding with the LAND-FRS
acceptance filter and providing the figures.
We thank to R.V. Jolos for his attention to the paper and the discussions.

\appendix
\section{The Rearrangement Potentials}\label{app:A}

Once the proton and neutron self-energies $\Sigma_{p,n}(\rho)$,
respectively, are known the rearrangements parts are determined and
properly subtracted by exploiting relations found in infinite
nuclear matter. In symmetric nuclear matter with $\rho_p=\rho_n$ we
find for the isoscalar self-energy $\Sigma_0=(\Sigma_n+\Sigma_p)/2$
the relation
\begin{equation}\label{eq:Sigma0}
\Sigma_0(\rho)=\frac{1}{2}\frac{\partial}{\partial \rho}\rho
U_0(\rho),
\end{equation}
which we integrate to give
\begin{equation}\label{eq:U0}
U_0(\rho)=\frac{2}{\rho}\int^\rho_0{d\rho'\Sigma_0(\rho')} \quad ,
\end{equation}
providing us with $U_0(\rho)=(U_n(\rho)+U_p(\rho))/2$. In pure
neutron matter we have $\rho=\rho_3=\rho_n$ and
\begin{equation}\label{eq:Un}
U_n(\rho)=\frac{2}{\rho}\int^\rho_0{d\rho'\Sigma_n(\rho')} \quad .
\end{equation}
This allows to determine
\begin{equation}\label{eq:Up}
U_p(\rho)=2U_0(\rho)-U_n(\rho) \quad .
\end{equation}
For a finite nucleus the densities are given parametrically as
functions of the radius $r$. Hence, we can replace the integrations
over density by radial integrals
\begin{equation}\label{eq:Uradial}
\rho(r) U_\alpha(r)= -2 \int_r^\infty{ds \frac{\partial
\rho(s)}{\partial s} \Sigma_\alpha(s)} \quad ,
\end{equation}
where $\rho(r)$ is the density calculated self-consistently
according to eq. \ref{eq:densities} with wave functions from the
effective potential $\Sigma_\alpha(r)$. Obviously, the above
equation is applicable to any potential given as a function of the
radius. Hence, by means of these {\em defolding relations} we are
able to calculate $B(A)$ for arbitrary phenomenological single-particle potentials, which otherwise we could not.


%

\end{document}